\documentclass[12pt, draftclsnofoot, onecolumn]{IEEEtran}

\usepackage{color}
\usepackage{cite}
\usepackage{url}

\ifCLASSINFOpdf
\else
\fi
\usepackage{amsfonts,amssymb}
\usepackage{latexsym}
\usepackage{graphicx}
\usepackage{epstopdf}
\usepackage{mathrsfs}
\usepackage{amsmath}
\usepackage{caption}
\usepackage{booktabs}
\usepackage{subfigure}
\usepackage{algorithm, algorithmic}
\usepackage{bbm}
\usepackage{bm}
\usepackage{caption}
\usepackage{amsmath}

\newtheorem{lemma}{\bf{Lemma}}

\usepackage{color}
\definecolor{myc1}{rgb}{0,0,0}
\definecolor{myc2}{rgb}{0,0,0}

\begin{document}


\title{Beamforming Design for the Performance Optimization of Intelligent Reflecting Surface Assisted Multicast MIMO Networks}

\author{{Songling Zhang,} \emph{Student Member, IEEE},  {Zhaohui Yang,} \emph{Member, IEEE},\\ {Mingzhe Chen,} \emph{Member, IEEE}, {Danpu Liu,} \emph{Senior Member, IEEE}, {Kai-Kit Wong,} \emph{Fellow, IEEE}, and H. Vincent Poor, \emph{Life Fellow, IEEE}\vspace*{-0.5em}\\ 
	\thanks{S. Zhang and D. Liu are with the Beijing Laboratory of Advanced Information Network, Beijing University of Posts and Telecommunications, Beijing, 100876, China (e-mail: \protect\url{slzhang@bupt.edu.cn}; \protect\url{dpliu@bupt.edu.cn)}.}
      \thanks{Z. Yang is with the College of Information Science and Electronic Engineering, Zhejiang University, Hangzhou 310027, China, and Zhejiang Provincial Key Lab of Information Processing, Communication and Networking (IPCAN), Hangzhou 310007, China, and also with Zhejiang Lab, Hangzhou 31121, China. (e-mail: \protect\url{yang_zhaohui@zju.edu.cn})}
	\thanks{M. Chen is with the Department of Electrical and Computer Engineering and Institute for Data Science and Computing, University of Miami, Coral Gables, FL, 33146 USA (e-mail: mingzhe.chen@miami.edu).}
 \thanks{K. K. Wong is with the Department of Electronic and Electrical Engineering, University College London, London, WC1E 6BT, UK (e-mail: \protect\url{kai-kit.wong@ucl.ac.uk}).}
	\thanks{H. V. Poor is with the Department of Electrical and
		Computer Engineering, Princeton University, Princeton, NJ, 08544, USA (e-mail: \protect\url{poor@princeton.edu}).}

}

\maketitle

\begin{abstract}
  In this paper, the problem of maximizing the sum of data rates of all users in an intelligent reflecting surface (IRS)-assisted millimeter wave multicast multiple-input multiple-output communication system is studied. In the considered model, one IRS is deployed to assist the communication from a multi-antenna base station (BS) to the multi-antenna users that are clustered into several groups. Our goal is to maximize the sum rate of  all users by jointly optimizing the transmit beamforming matrices of the BS, the receive beamforming matrices of the users, and the phase shifts of the IRS. To solve this non-convex problem, we first use a block diagonalization method to represent the beamforming matrices of the BS and the users by the phase shifts of the IRS. Then, substituting the expressions of the beamforming matrices of the BS and the users, the original sum-rate maximization problem can be transformed into a problem that only needs to optimize the phase shifts of the IRS. To solve the transformed problem, a manifold method is used.  Simulation results show that the proposed scheme can achieve up to 28.6\%  gain in terms of  the sum rate  of all users compared to the algorithm that optimizes the hybrid beamforming matrices of the BS and the users using our proposed scheme and randomly determines the phase shifts of the IRS.

\end{abstract}


\section{Introduction}

Millimeter wave (mmWave) communications, which utilize the 30-300 GHz frequency band to achieve multi-gigabit data rates,  is a promising technology for emerging and envisioned wireless  systems\cite{6894453,9813455,9433528,9795144}. However, mmWave suffers from severe path loss and is easily blocked by obstacles due to the short wavelengths\cite{6515173}. To address these problems, massive multiple-input multiple-output (MIMO) and intelligent reflecting surfaces (IRSs) have been proposed \cite{9770804, 8371237,9497709,8741198,9127834}. However, deploying IRSs and massive MIMO in mmWave communication systems faces several challenges such as IRS deployment optimization, and joint active and passive beamforming design.

Recently, a number of  works 
 have studied important problems related to the deployment of IRSs in wireless networks.  The work in \cite{9777870}  considered the maximization of the spectral
efficiency by separately designing the passive beamforming matrix and
active precoder.  The authors in \cite{9690059} jointly designed a hybrid precoder at a base station (BS) and a passive precoder
at the IRS to maximize the average spectral efficiency in an
IRS-assisted mmWave MIMO system.    To maximize the end-to-end signal-to-noise ratio (SNR), the authors in \cite{9324795} optimized the phase shifts of the IRS.            The work
in  \cite{9226616} maximized the received signal power by jointly optimizing a transmit precoding vector of the base station (BS) and the phase shift coefficients of an IRS. The authors in \cite{9234098} maximized the spectral efficiency by jointly optimizing the  reflection coefficients of the IRS, a hybrid precoder at the BS and a hybrid combiner at the end-user device. The work in  \cite{9032163} studied  hybrid precoding design for an IRS aided multi-user mmWave  communication system.   In \cite{8964330}, a geometric mean decomposition-based beamforming scheme was  proposed for IRS-assisted mmWave hybrid MIMO systems.  In \cite{9214532},  the authors optimized a channel estimator in  closed form while considering the signal reflection matrix of an IRS and an analog combiner at the receiver. The authors in \cite{9246254} jointly optimized the  transmit beamforming vectors of  multiple BSs and the reflective beamforming vector of the IRS so as to maximize the minimum weighted signal-to-interference-plus-noise ratio (SINR) of users.
In \cite{9735360},  the authors studied
the deployment of multiple IRSs to improve the spatial diversity gain and designed a robust beamforming scheme based on stochastic optimization methods to minimize the maximum outage probability among multiple users.  The work in  \cite{9819898} jointly designed a
hybrid precoder at the BS and the passive precoders at
the IRSs to maximize the spectral efficiency.       The authors in \cite{9629293}  investigated the use of double IRSs to improve the spectral efficiency in a
multi-user MIMO  network operating
in the mmWave band.     The work in  \cite{9362274} studied a double IRS assisted multi-user communication system with a cooperative passive beamforming design. However, most of these existing works \cite{9777870,9690059,9324795,9226616,9234098,9032163,8964330,9214532,9246254,9735360,9819898,9629293,9362274} only consider the deployment of IRSs over unicast communication networks in which each BS transmits independent data streams to different users.

Multicast enables the BS to transmit  content to multiple users using one radio resource block, thus improving the spectral and energy efficiency \cite{8964475,9685146,7937794}. However, deploying IRSs over multicast communication systems faces several new challenges. First, the users in a group that have different channel conditions need to be served by a coordinated  beamforming matrix, thus complicating the design of the beamforming matrix of the transmitter. Moreover, in a multicast system, the data rate of a group is limited by the user with the worst channel gain. Therefore, in a multicast system, one must maximize the data rate of the user with the worst channel gain in each group.

Several existing studies \cite{9186127,9076830,9435051,9431722} have considered the use of IRSs in multicast communication systems. In particular, the work in \cite{9186127} studied a multicast system where a
single-antenna transmitter sends a common message to multiple
single-antenna users via an IRS.  In \cite{9076830},  the authors  maximized the sum rate of all multicasting
groups by the joint optimization of the precoding matrix
at the base station  and the reflection coefficients at the IRS
under both  power and unit-modulus constraints.   The authors in\cite{9435051} considered an IRS assisted multicast transmission scenario, where a base
station (BS) with multiple-antenna multicasts common message to
multiple single-antenna  users  under the assistance
of an IRS.   The work in \cite{9431722} optimized the energy efficiency of
an IRS-assisted multicast communication network.  However, these existing works \cite{9186127,9076830,9435051,9431722} neither considered  mmWave nor the use of hybrid beamforming  at the BS and the users.  Considering mmWave and the use of hybrid beamforming  at the BS and the users in an IRS-assisted multicast communication system faces several challenges such as severe path loss, joint analog and digital precoder design and optimization for a BS that uses mmWave and multicast techniques to serve users, and jointly optimizing the transmit beamforming matrices of the BS, the receive beamforming matrices of the users, and the phase shifts of the IRS.

The main contribution of this paper is to develop a novel IRS assisted multigroup multicast MIMO system. To the best of our knowledge, this is the first work that studies the joint use of an IRS, the hybrid beamforming  at the BS and the users, the mmWave band, multicast, and MIMO to service the users in several groups. The key contributions are summarized as follows: 
\begin{itemize}
	
\item	
We consider an IRS-assisted mmWave multicast MIMO communication system. In the considered model, one IRS is used to assist the communication from a multi-antenna BS to multi-antenna users that are clustered into several groups. To maximize the sum rate of all the multicasting groups, we jointly optimize the transmit beamforming matrices of the BS, the receive beamforming matrices of the users, and the phase shifts of the IRS. We formulate an optimization problem with the objective of maximizing the sum rate of all the multicasting groups under amplitude constraints on radio frequency (RF)  beamforming matrices, maximum transmit power constraint, and unit-modulus constraint of the IRS phase shifts. 

\item
To solve this problem, we first use a block diagonalization (BD) method to represent the beamforming matrices of the BS and the users in terms of  the phase shifts of the IRS. Then, we substitute the expressions of the beamforming matrices of the BS and the users into the original problem so as to transform it to a problem that only needs to optimize the phase shifts of the IRS. The transformed  problem is then solved by a manifold method.

\end{itemize}
Simulation results show that the proposed scheme can achieve up to 28.6\% gain in terms of  the sum rate of all the multicasting groups compared to the algorithm that optimizes the hybrid beamforming matrices of the BS and the users using our proposed algorithm and randomly determines the phase shifts  of the IRS.

The rest of this paper is organized as follows. The system
model and problem formulation are described in Section II.
The algorithm is introduced in Section III.  Simulation results
are presented in Section IV. Conclusions are drawn in Section V.

\begin{table}[htbp]
	\renewcommand{\arraystretch}{1.3}
	\begin{center}
		{\color{myc2}{    \caption{List of Main Notation.}}}
		\label{tab:table1}
		\resizebox{\textwidth}{!}{
		\begin{tabular}{|c||c|c||c|}   \hline
			\textbf{Notation} & \textbf{Description} & \textbf{Notation} & \textbf{Description}\\
			\hline
			$N^\textrm{B}$  & Number of antennas of the BS &$K$ & Number of users     \\   \hline
			$H$ & Number of user groups &  $\mathcal {H} $ &The set of user groups     \\   \hline
			$N^\textrm{U}$ & Number of antennas of each user&  $M^\textrm{U}$& Number of RF chains of each user   \\   \hline
			$\zeta$ & Data streams received by each user&  $M^\textrm{B}$ & Number of RF chains of the BS      \\   \hline
			$\boldsymbol{F}^\textrm{B}$ & Baseband transmit beamforming matrix &  $\boldsymbol{F}^\textrm{B}_{h}$ & Transmit beamforming matrix of group $h$   \\   \hline
			$\boldsymbol{F}^\textrm{R}$ & RF transmit beamforming matrix &  $\boldsymbol{W}_{k}^{\textrm{R}} $ & RF receive beamforming matrix of user $k$     \\   \hline
		$\boldsymbol{W}_{k}^{\textrm{B}}$ & Baseband receive beamfoming matrix of user $k$  &  $\boldsymbol{\Phi}  $ & Phase-shift matrix of the IRS    \\   \hline
			$M$ & Number of reflecting elements at the IRS & ${\phi _m}  $ & Phase shift introduced by element $m$ of the IRS    \\   \hline
			$N$ & Number of antennas in ULA &   $d$ & Interval between two antennas   \\   \hline
			$\lambda $ & Signal wavelength &   ${F_y}$ & Number of elements in the horizontal  directions    \\   \hline
			${F_z}$ & Number of elements in the vertical directions & $\boldsymbol{H}^\textrm{B} $ & BS-IRS channel     \\   \hline
			$\boldsymbol{H}^\textrm{R}_{k}$ &  Channel  from the IRS to user $k$  &   $Y$  &  Total number of paths between the BS and the IRS \\   \hline
			$\theta _i^{\textrm{A}}$ &  Azimuth angle of arrival of the IRS & $L$  &  Total number of paths between the IRS and user $k$ \\   \hline
			$\theta _i^{\textrm{D}}$ & Azimuth angle of departure of the IRS & $\eta _i^{\textrm{A}}$ &  Elevation angle of arrival of the IRS  \\   \hline
			$\eta _i^{\textrm{D}}$ & Elevation angle of departure of the IRS &$r_{i,k}^{\textrm{A}}$ & Arrival angle of user $k$    \\   \hline
		$r_i^{\textrm{D}}$ & Departure angle of the BS &  $\hat{\boldsymbol{s}}_{k,h}$  &Detected data of user $k$ in group $h$   \\   \hline
			$\boldsymbol{s}$ & Data streams to be transmitted to all users  &  ${\boldsymbol{a}}\left(r_i^{\textrm{D}}\right)$ & Normalized array response vectors of the BS  \\   \hline
			$\boldsymbol{a}\left(r_{i,k}^{\textrm{A}}\right) $ & Normalized array response vectors of the  user $k$ &  ${\boldsymbol{H}}_{k} $ & Effective  channel from the BS to user $k$     \\   \hline
			$\boldsymbol{s}_h$ & $\zeta$ streams to be transmitted to each user in group $h$  &  $\boldsymbol{n}_{k}$ & Additive white Gaussian noise vector of user $k$  \\   \hline
		$\hat{s}_{ik,h}$ & Estimated data stream $i$ received by user $k$ in group $h$  &  $\xi _{ik,h} $ & SINR of user $k$ in group $h$ receiving data stream $i$     \\   \hline
			${I_{ik,h}}$& Interference from user itself&  ${J_{ik,h}}$ & Interference from other groups    \\   \hline
			${R_{k,h}}$ & Achievable data rate of user $k$ in group $h$ & $P$ &  transmit power of the BS     \\   \hline
			$\boldsymbol{B}$ & Fully digital transmit beamforming matrix&  $\boldsymbol{J}_{k} $& Fully digital receive beamforming matrix of user $k$ in group $h$   \\   \hline
			$p_i$ &Transmit power of data stream $i$ &   ${{\boldsymbol{G}}_{h}}   $ & Power allocation matrix  in group $h$   \\   \hline
			$\nabla f\left({{\boldsymbol{\nu}}_n}\right) $ & Euclidean gradient & $\mathcal Q$& Oblique manifold    \\   \hline
		${{{ T}_{{\boldsymbol{\nu}}_{n}}}\mathcal Q}$ & Tangent space  &   ${\mathcal G _{{\boldsymbol{\nu}}_{n}} \mathcal Q}$ &  Riemannian gradient  \\   \hline
		\end{tabular}}
	\end{center}
\end{table}

\section{System Model And Problem Formulation}\label{SYSTEM MODEL AND PROBLEM FORMULATIO}

\subsection{System Model}\label{S1}
\begin{figure}[t]
	\centering
	\includegraphics[width=7.8cm]{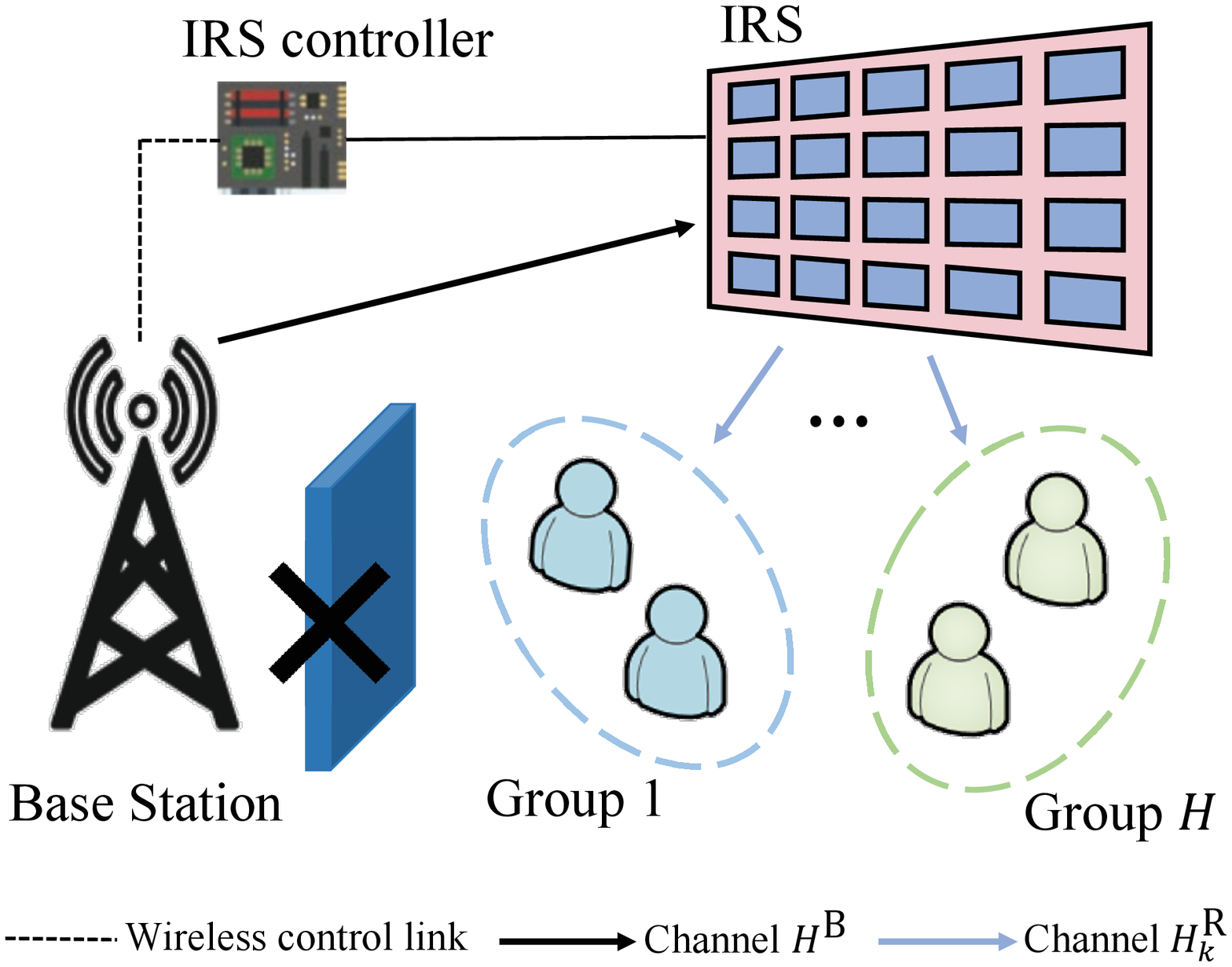}
	\caption{An IRS-aided mmWave multigroup multicast MIMO communication system.}\label{fig 1}
	\vspace{-0.5cm}
\end{figure}
We consider an IRS-aided mmWave multigroup multicast MIMO communication system in which a BS is equipped with $N^\textrm{B}$ antennas serving $K$ users via an IRS, as shown in Fig. \ref{fig 1}.
The users are divided into $H$ groups. We assume that the users in a group will request the same data streams and the data streams requested by the users in different groups are different. The set of user groups is denoted by $\mathcal {H}= \left\{ 1,2,\ldots,H \right\}$. Meanwhile, the set of users in a group $h$ is denoted as $\mathcal H_h$. We also assume that each user can only belong to one group,  i.e., $\mathcal H_i \cap \mathcal H_j=\emptyset$, $\forall i,j \in \mathcal H,i \ne j$. In our model, the direct communication link between the BS and a user is blocked due to unfavorable propagation conditions. Each user is equipped with $N^\textrm{U}$ antennas and $M^\textrm{U}$ RF chains to receive $\zeta$ data streams from the BS. The BS simultaneously transmits $H\zeta$ independent data streams to the users by $M^\textrm{B}$ RF chains\footnote{The numbers of RF chains are subject to the constraints $ H\zeta \le M^\textrm{B} \le N^\textrm{B}$ and $ \zeta \le M^\textrm{U} \le N^\textrm{U}$.}. {\color{myc2}{
		The main notations used in this work are summarized in Table I.}}

At the BS, the transmitted data streams of $H$ user groups  are precoded by a baseband transmit beamforming matrix $\boldsymbol{F}^\textrm{B} = \left[\boldsymbol{F}^\textrm{B}_{1}, \boldsymbol{F}^\textrm{B}_{2},\ldots,\boldsymbol{F}^\textrm{B}_{H} \right] \in \mathbb{C}^{M^\textrm{B} \times H\zeta}$, with $\boldsymbol{F}^\textrm{B}_{h}$ being the transmit beamforming matrix of group $h$. After that, each transmitted data stream of $H$ user groups is precoded by an RF transmit beamforming matrix $\boldsymbol{F}^\textrm{R} \in \mathbb{C}^{N^\textrm{B} \times M^\textrm{B}}$.

The received data streams of user $k$ in group $h$ are first processed by an RF receive beamforming matrix $\boldsymbol{W}_{k}^{\textrm{R}} \in \mathbb{C}^{N^\textrm{U} \times M^\textrm{U}}$. Then, user $k$ uses a baseband receive beamfoming matrix $\boldsymbol{W}_{k}^{\textrm{B}} \in \mathbb{C}^{M^\textrm{U} \times \zeta}$ to recover $\zeta$ data streams.

In our model, an IRS is used to enhance the received signal strength of users by reflecting signals from the BS to the users. We assume that the signal power of the multi-reflections (i.e., reflections more than once) on the IRS is ignored due to severe path loss. The phase-shift matrix of the IRS is $\boldsymbol{\Phi}  = \text{diag}\left({e^{j{\phi _1}}},\ldots,{e^{j{\phi _M}}}\right) \in \mathbb{C}^{M \times M}$,  where $\text{diag}\left({e^{j{\phi _1}}},\ldots,{e^{j{\phi _M}}}\right)$ is a diagonal matrix of $\left[{e^{j{\phi _1}}},\ldots,{e^{j{\phi _M}}}\right]$, $M$ is the number of reflecting elements at the IRS, and ${\phi _m} \in \left[0,2\pi \right]$ is the phase shift introduced by element $m$ of the IRS.

\subsubsection{Channel Model}
The BS and the users employ uniform linear arrays (ULAs), and the IRS  uses a uniform planar array (UPA). The normalized array response vector for an ULA is
\begin{equation}\label{System model}
\boldsymbol{a}\left(r\right) = \frac{1}{{\sqrt N }}{\left[1, \cdots, {e^{j\frac{{2\pi d}}{\lambda }\left(n - 1\right)\sin \left(r\right)}}, \cdots, {e^{j\frac{{2\pi d}}{\lambda }\left(N - 1\right)\sin \left(r\right)}}\right]^{\textrm{T}}},
\end{equation}
where $N$ is the number of antennas in ULA, $d$ is an interval between two antennas, and $\lambda $ is the signal wavelength. The normalized array response vector of UPA is
\begin{equation}\label{222222}
\begin{split}
\boldsymbol{a}\left(\theta ,\eta \right) = &\frac{1}{{\sqrt {{F_y} \times {F_z}} }}[1, \cdots, {e^{j\frac{{2\pi d}}{\lambda }\left(\left({f_1} - 1\right)\cos \left(\eta \right)\sin \left(\theta \right) + \left({f_2} - 1\right)\sin \left(\eta \right)\right)}},\\
&{\rm{                     }} \cdots, {e^{j\frac{{2\pi d}}{\lambda }\left(\left({F_y} - 1\right)\cos \left(\eta \right)\sin \left(\theta \right) + \left({F_z} - 1\right)\sin \left(\eta \right)\right)}}]^{\textrm{T}},
\end{split}
\end{equation}
where ${F_y} \times {F_z}$ is the number of elements in UPA, $F_y$ and $F_z$ are respectively the number of elements in the horizontal and vertical directions. The BS-IRS channel  $\boldsymbol{H}^\textrm{B}\in \mathbb{C}^{M \times N^\textrm{B}}$ and the channel  $\boldsymbol{H}^\textrm{R}_{k} \in \mathbb{C}^{N^\textrm{U} \times M}$ from the IRS to user $k$ in group $h$ can be respectively given as
\begin{equation}\label{System model111}
\boldsymbol{H}^\textrm{B}= \sqrt {\frac{{{N^\textrm{B}}M}}{Y}} \sum\limits_{i = 1}^Y {\alpha _i} \boldsymbol{a}\left(\theta _i^{\textrm{A}},\eta _i^{\textrm{A}}\right) \left(\boldsymbol{a}\left(r_i^{\textrm{D}}\right) \right)^{\textrm{H}},
\end{equation}
\begin{equation}\label{System model222}
\boldsymbol{H}^\textrm{R}_{k} = \sqrt {\frac{{M{N^\textrm{U}}}}{L}} \sum\limits_{i = 1}^L {\beta _i}\boldsymbol{a}\left(r_{i,k}^{\textrm{A}}\right) \left(\boldsymbol{a}\left(\theta _i^{\textrm{D}},\eta _i^{\textrm{D}}\right)\right)^{{\textrm{H}}},
\end{equation}
where $Y$ is the total number of paths (line-of-sight (LOS) and non-line-of-sight (NLOS)) between the BS and the IRS, $L$ is the total number of paths (LOS and NLOS) between the IRS and user $k$, $\theta _i^{\textrm{A}}$ denotes  the azimuth angle of arrival of the IRS, $\theta _i^{\textrm{D}}$ denotes the azimuth angle of departure of the IRS, $\eta _i^{\textrm{A}}$ denotes the elevation angle of arrival of the IRS,   $\eta _i^{\textrm{D}}$ denotes the elevation angle of departure of the IRS, $r_{i,k}^{\textrm{A}}$ represents the arrival angle of user $k$, $r_i^{\textrm{D}}$ represents the  departure angle of the BS, ${\alpha _i}$ and ${\beta _i}$  are  complex channel gains. ${\boldsymbol{a}}\left(r_i^{\textrm{D}}\right)$ and $\boldsymbol{a}\left(r_{i,k}^{\textrm{A}}\right) $ denote the normalized array response vectors of the BS and user $k$, respectively. $\left(\boldsymbol{a}\left(r_i^{\textrm{D}}\right) \right)^{\textrm{H}}$ is the Hermitian transpose of ${\boldsymbol{a}}\left(r_i^{\textrm{D}}\right)$. $\boldsymbol{a}\left(\theta _i^{\textrm{A}},\eta _i^{\textrm{A}}\right)$ represents the normalized array response vector of the IRS over the effective channel from the BS to the IRS. $\boldsymbol{a}\left(\theta _i^{\textrm{D}},\eta _i^{\textrm{D}}\right)$ represents the normalized array response vector of the IRS over the effective channel from the IRS to user $k$. The effective  channel from the BS to user $k$ in group $h$ is ${\boldsymbol{H}}_{k} = {G_\textrm{t}} {G_\textrm{r}} \boldsymbol{H}^\textrm{R}_{k} \boldsymbol{\Phi} \boldsymbol{H}^\textrm{B}$, where ${G_\textrm{t}}$ and ${G_\textrm{r}}$ are the antenna gains of the BS and each user, respectively.

\subsubsection{Data Rate Model}
We assume that the BS obtains the channel state information (CSI). The BS is responsible for designing the reflection coefficients of the IRS. As a result, the detected data of user $k$ in group $h$ is given by
\begin{equation}\label{System model1}
\hat{\boldsymbol{s}}_{k,h}\! =\! \left(\boldsymbol{W}_{k}^{\textrm{B}}\right)^{\textrm{H}} \left(\boldsymbol{W}_{k}^{\textrm{R}}\right)^{\textrm{H}} \boldsymbol{H}_{k} \boldsymbol{F}^\textrm{R} \boldsymbol{F}^\textrm{B} \boldsymbol{s} + \left(\boldsymbol{W}_{k}^{\textrm{B}}\right)^{\textrm{H}} \left(\boldsymbol{W}_{k}^{\textrm{R}}\right)^{\textrm{H}} \boldsymbol{n}_{k},
\end{equation}
where $\boldsymbol{s} = \left [ \boldsymbol{s}_1^{\textrm{T}} , \ldots , \boldsymbol{s}_H^{\textrm{T}} \right ]^{\textrm{T}} \in \mathbb{C}^{H\zeta \times 1}$ represents the data streams to be transmitted to all users, with $\boldsymbol{s}_h=\left[s_{h,1},\ldots, s_{h,\zeta}\right]^{\textrm{T}} \in \mathbb{C}^{\zeta \times 1}$ being $\zeta$ streams that will be transmitted to each user in group $h$. $\boldsymbol{n}_{k} \in \mathbb{C}^{N^{\textrm{U}} \times 1}$ is an additive white Gaussian noise vector of user $k$. Each element of $\boldsymbol{n}_{k}$ follows the independent and identically distributed complex Gaussian distribution with zero mean and variance $\sigma^2$. In (\ref{System model1}), the first term represents the signal received by user $k$. The second term is the noise received by user $k$. The estimated data stream $i$ received by user $k$ in group $h$ can be expressed as
\begin{equation}\label{System model2}
\begin{split}
\hat{s}_{ik,h} &= \left(\boldsymbol{w}_{k,i}^{\textrm{B}}\right)^{\textrm{H}} \left(\boldsymbol{W}_{k}^{\textrm{R}}\right)^{\textrm{H}} \boldsymbol{H}_{k} \boldsymbol{F}^\textrm{R} \bar{\boldsymbol{f}}_{h_i}^{\textrm{B}}  s_{h,i} \\
&+ \sum_{j=1,j \neq i}^{\zeta} \left(\boldsymbol{w}_{k,i}^{\textrm{B}}\right)^{\textrm{H}} \left(\boldsymbol{W}_{k}^{\textrm{R}}\right)^{\textrm{H}} \boldsymbol{H}_{k} \boldsymbol{F}^\textrm{R} \bar{\boldsymbol{f}}_{h_j}^{\textrm{B}} s_{h,i} \\
&+ \sum_{m=1,m  \notin  {\mathcal H_h}}^{H} \sum_{l=1}^{\zeta} \left(\boldsymbol{w}_{k,i}^{\textrm{B}}\right)^{\textrm{H}} \left(\boldsymbol{W}_{k}^{\textrm{R}}\right)^{\textrm{H}} \boldsymbol{H}_{k} \boldsymbol{F}^\textrm{R} \bar{\boldsymbol{f}}_{m_l}^{\textrm{B}}  s_{m,l} \\
&+ \left(\boldsymbol{w}_{k,i}^{\textrm{B}}\right)^{\textrm{H}} \left(\boldsymbol{W}_{k}^{\textrm{R}}\right)^{\textrm{H}} \boldsymbol{n}_{k},
\end{split}
\end{equation}
where $h_i = \left(h-1\right)\zeta+i$, $\boldsymbol{w}_{k,i}^{\textrm{B}}$ denotes row $i$ of matrix $\boldsymbol{W}_{k}^{\textrm{B}}$,  and $\bar{\boldsymbol{f}}_{h_i}^{\textrm{B}}$ denotes column $h_i$ of matrix $\boldsymbol{F}^\textrm{B}$.     In (\ref{System model2}), the first term represents the desired signal. The second term is the interference caused by other streams of user $k$. The third term is the  interference caused by the users from other groups. The fourth term is the noise.
The SINR of user $k$ in group $h$ receiving data stream $i$ is given by
\begin{equation}\label{R-}
\begin{split}
{{\xi _{ik,h}\left(\boldsymbol{W}_{k}^{\textrm{R}}, \boldsymbol{W}_{k}^{\textrm{B}}, {\boldsymbol{\nu}},\boldsymbol{F}^\textrm{R}, \boldsymbol{F}^\textrm{B}_{h}\right)} }&= \\
&\!\!\!\!\!\!\!\!\!\!\!\!\!\! \frac{\left | \left(\boldsymbol{w}_{k,i}^{\textrm{B}}\right)^{\textrm{H}} \left(\boldsymbol{W}_{k}^{\textrm{R}}\right)^{\textrm{H}} \boldsymbol{H}_{k} \boldsymbol{F}^\textrm{R} \bar{\boldsymbol{f}}_{h_i}^{\textrm{B}} \right |^2}        {{I_{ik,h}}+{J_{ik,h}}+\sigma^2},
\end{split}
\end{equation}
where ${I_{ik,h}}$ is short for ${I_{ik,h}}\left(\boldsymbol{W}_{k}^{\textrm{R}}, \boldsymbol{W}_{k}^{\textrm{B}}, {\boldsymbol{\nu}},\boldsymbol{F}^\textrm{R}, \boldsymbol{F}^\textrm{B}_{h}\right)$ and ${I_{ik,h}}= \sum_{j=1,j \neq i}^{\zeta} \left | \left(\boldsymbol{w}_{k,i}^{\textrm{B}}\right)^{\textrm{H}} \left(\boldsymbol{W}_{k}^{\textrm{R}}\right)^{\textrm{H}} \boldsymbol{H}_{k} \boldsymbol{F}^\textrm{R} \bar{\boldsymbol{f}}_{h_j}^{\textrm{B}} \right |^2$ represents the interference from user itself,   ${J_{ik,h}}$ is short for ${J_{ik,h}}\left(\boldsymbol{W}_{k}^{\textrm{R}}, \boldsymbol{W}_{k}^{\textrm{B}}, {\boldsymbol{\nu}},\boldsymbol{F}^\textrm{R}, \boldsymbol{F}^\textrm{B}_{h}\right)$ and $J_{ik,h}= \sum_{m=1,m  \notin  {\mathcal H_h}}^{H} \sum_{l=1}^{\zeta} \left | \left(\boldsymbol{w}_{k,i}^{\textrm{B}}\right)^{\textrm{H}} \left(\boldsymbol{W}_{k}^{\textrm{R}}\right)^{\textrm{H}} \boldsymbol{H}_{k} \boldsymbol{F}^\textrm{R} \bar{\boldsymbol{f}}_{m_l}^{\textrm{B}}\right |^2$  represents the interference from other groups. The achievable data rate of user $k$ in group $h$ is given by
\begin{equation}\label{R}
\begin{split}
&{R_{k,h}}\left(\boldsymbol{W}_{k}^{\textrm{R}}, \boldsymbol{W}_{k}^{\textrm{B}},{\boldsymbol{\nu}},\boldsymbol{F}^\textrm{R}, \boldsymbol{F}^\textrm{B}_{h}\right)= \\
&\;\;\;\;\;\;\;\;\;\;\, W\sum_{i=1}^{\zeta} \log_2 \left ( 1 + {\xi _{ik,h}\left(\boldsymbol{W}_{k}^{\textrm{R}}, \boldsymbol{W}_{k}^{\textrm{B}},{\boldsymbol{\nu}},\boldsymbol{F}^\textrm{R}, \boldsymbol{F}^\textrm{B}_{h}\right)} \right ),
\end{split}
\end{equation}
where $W$ is the bandwidth.

Due to the nature of the multicast mechanism, the achievable data rate of group $h$ depends on the user with minimum data rate, which is defined as follows:
\begin{equation}\label{System model}
\mathop {\min }\limits_{k \in {\mathcal H_h}} \left\{{R_{k,h}}\left(\boldsymbol{W}_{k}^{\textrm{R}}, \boldsymbol{W}_{k}^{\textrm{B}},{\boldsymbol{\nu}},\boldsymbol{F}^\textrm{R}, \boldsymbol{F}^\textrm{B}_{h}\right) \right\}.
\end{equation}

\subsection{Problem Formulation}\label{S123}
Next, we introduce our optimziation problem. Our goal is to maximize the sum rate of all the multicasting groups via jointly optimizing the transmit beamforming matrices $\boldsymbol{F}^\textrm{B}$, $\boldsymbol{F}^\textrm{R}$, the receive beamforming matrices $\boldsymbol{W}^\textrm{R}$, $\boldsymbol{W}^\textrm{B}$, and the phase shift ${\boldsymbol{\nu}}$ of the IRS. Mathematically, the optimization problem is formulated as

\begin{subequations}   \label{PF1}
	\begin{align}
	 &\mathop {\max}_{\boldsymbol{W}_{k}^\textrm{B},\boldsymbol{W}_{k}^\textrm{R},\boldsymbol{F}^\textrm{R},\boldsymbol{F}^\textrm{B},{\boldsymbol{\nu}} }     \sum\limits_{h = 1}^H {\mathop {\min }\limits_{k \in {\mathcal H_h}} \!\!\left\{ {{R_{k,h}}\left(\boldsymbol{W}_{k}^{\textrm{R}}, \boldsymbol{W}_{k}^{\textrm{B}},{\boldsymbol{\nu}},\boldsymbol{F}^\textrm{R}, \boldsymbol{F}^\textrm{B}_{h}\right)}\right\} } \tag{\theequation} \\
	&\textrm{s.t.}  \ \ \left\|\boldsymbol{F}^\textrm{R}\boldsymbol{F}^\textrm{B}\right\|_F^2 \le P,\\
    &\quad\: \ \ {\rm{      }}\left| \boldsymbol{F}^\textrm{R}\left(i,j\right) \right| = \left| \boldsymbol{W}_{k}^{\textrm{R}}\left(i,j\right) \right| = 1, \forall i,j,\\
	 & 	\quad\ \ \   0\le \phi _m \le 2\pi, m=1,\ldots,M,
	\end{align}
\end{subequations}
where $P$ is the  transmit power of the BS,  $\left\|\boldsymbol{F}^\textrm{R}\boldsymbol{F}^\textrm{B}\right\|_F$ is the Frobenius norm of $\boldsymbol{F}^\textrm{R}\boldsymbol{F}^\textrm{B}$, $\boldsymbol{\nu}  = {\left[{e^{j{\phi _1}}}, \ldots, {e^{j{\phi _M}}}\right]^{\textrm{H}}}$, $ \boldsymbol{F}^\textrm{R}\left(i,j\right) $ denotes the element $\left(i,j\right)$ of matrix $\boldsymbol{F}^\textrm{R}$, with $\left| \boldsymbol{F}^\textrm{R}\left(i,j\right) \right|$ being the amplitude of $ \boldsymbol{F}^\textrm{R}\left(i,j\right) $. The  transmit power constraint of the BS is given in (\ref{PF1}a). Constraint (\ref{PF1}b) represents the  amplitude constraints of the RF beamforming matrices of the BS and each user, while (\ref{PF1}c) shows the phase shift limits of the IRS. Due to non-convex objective function (\ref{PF1}) and non-convex constraints (\ref{PF1}a)-(\ref{PF1}c), problem (\ref{PF1}) is non-convex, and hence it is hard to solve. Next, we introduce an efficient scheme to solve problem (\ref{PF1}).
\section{Proposed  Scheme}
In this section, we first use the phase shift ${\boldsymbol{\nu}}$ of the IRS to represent  the fully digital transmit beamforming matrix of the BS and receive beamforming matrix of the users. Then, we substitute them in (\ref{PF1}) to transform problem (\ref{PF1}). To solve the transformed problem, the phase shift ${\boldsymbol{\nu}}$ of the IRS is optimized by a manifold method. Finally, we introduce the entire algorithm used to solve problem (\ref{PF1}).

\subsection{Block Diagonalization Method}

\subsubsection{Simplification of Optimization Problem}
$\boldsymbol{W}_{k}^\textrm{B}\boldsymbol{W}_{k}^\textrm{R}$ and $\boldsymbol{F}^\textrm{R}\boldsymbol{F}^\textrm{B}$ are regarded as a whole, that is, problem (\ref{PF1}) is solved as a problem of fully digital beamforming. Once the fully digital transmit beamforming matrix  and receive beamforming matrix are obtained, we can use the algorithm in \cite{8646553} to find  the hybrid transmit beamforming matrices and receive beamforming matrices to approximate the fully digital  transmit beamforming matrix and receive beamforming matrix, as done in \cite{7397861,6717211}.  Let $\boldsymbol{B} = \left[\boldsymbol{B}_1,\ldots,\boldsymbol{B}_H \right] \in \mathbb{C}^{N^\textrm{B} \times H\zeta}$ be a fully digital transmit beamforming matrix which has the same size as the hybrid transmit beamforming matrix $\boldsymbol{F}^\textrm{R}\boldsymbol{F}^\textrm{B}$ and $\boldsymbol{J}_{k} \in \mathbb{C}^{N^\textrm{U} \times \zeta}$ be a fully digital receive beamforming matrix of user $k$ in group $h$. The size of $\boldsymbol{J}_{k}$ is equal to that of the hybrid receive beamforming matrix $\boldsymbol{W}_{k}^{\textrm{R}} \boldsymbol{W}_{k}^{\textrm{B}}$. Substituting $\boldsymbol{B}$ and $\boldsymbol{J}_{k}$ in (\ref{PF1}), problem (\ref{PF1}) can be transformed as
\begin{subequations}\label{PF2}
\begin{align}
\mathop {\max }_{\boldsymbol{B}, \boldsymbol{J},  {\boldsymbol{\nu}} }  & \quad  \sum\limits_{h = 1}^H {\mathop {\min }\limits_{k \in {\mathcal H_h}} \left\{ {{R_{k,h}}\left({{\boldsymbol{B}}_h,  {\boldsymbol{J}}_{k}, {\boldsymbol{\nu}} }\right)}\right\} } \tag{\theequation} \\
\textrm{s.t.} \quad &   \left(\ref{PF1}\textrm{c}\right), \notag\\
&\, \left\|\boldsymbol{B}\right\|_{\textrm{F}}^2 \le P,
\end{align}
\end{subequations}
where $\boldsymbol{J} = \text{diag} \left(\boldsymbol{J}_{1},\ldots,\boldsymbol{J}_{K} \right) $.

\subsubsection{Optimization of $\boldsymbol{B}$ and $\boldsymbol{J}$ }
Due to the low complexity of a BD method, we use it to find the relationship between ${\boldsymbol{\nu}}$ and the fully digital transmit beamforming matrix ${\boldsymbol{B}}$ of the BS as well as the receive beamforming matrix ${\boldsymbol{J}}$ of the users.

\begin{lemma}\label{thm1}
	Given ${\boldsymbol{\nu}}$ and the power allocation matrix ${{\boldsymbol{G}}_{h}} = \text{diag}\left(p_1,\ldots,p_{\zeta}\right)$ in group $h$, where $p_i={{\frac{P}{{H\zeta}}}}$ is the transmit power of data stream $i$, $\boldsymbol{B}$ and $\boldsymbol{J}$ can be given by
	\begin{equation}\label{PF8}
{\boldsymbol{B}}_h\left({\boldsymbol{\nu}}\right) = \tilde{\boldsymbol{V}}_{h}^{\left(0\right)} \left(\frac{{\boldsymbol{V}}_1^{\left(1\right)}+\cdots+ {\boldsymbol{V}}_K^{\left(1\right)} }{\sqrt{\left | \mathcal H_h \right |}}\right) {\sqrt{\frac{P}{{H\zeta}}}},
	\end{equation}

	\begin{equation}\label{PF7+}
	{\boldsymbol{J}}_{k}\left({\boldsymbol{\nu}}\right)={\boldsymbol{U}}_k^{\left(1\right)},
	\end{equation}
\end{lemma}
where $ \tilde{\boldsymbol{V}}_{h}^{\left(0\right)}=\text{null} \left(\tilde{\boldsymbol{H}}_{h}\right)$, ${\boldsymbol{U}}_k^{\left(1\right)}$, and ${{\boldsymbol{V}}_{k}^{\left(1\right)} }$ can be obtained by singular value decomposition (SVD) of ${\boldsymbol{H}}_{k} \tilde{\boldsymbol{V}}_{h}^{\left(0\right)}$ with ${\boldsymbol{H}}_{k} \tilde{\boldsymbol{V}}_{h}^{\left(0\right)}= \left [ {\boldsymbol{U}}_{k}^{\left(1\right)} , {\boldsymbol{U}}_{k}^{\left(2\right)} \right ] \left [ \begin{matrix}  {\boldsymbol{\Sigma}}_{k}^{\left(1\right)}&\boldsymbol{0}&\\\boldsymbol{0}&{\boldsymbol{\Sigma}}_{k}^{\left(2\right)}& \end{matrix} \!\!\!\!\! \right ] \left [ {\boldsymbol{V}}_{k}^{\left(1\right)} , {\boldsymbol{V}}_{k}^{\left(2\right)} \right ]^{\textrm{H}}$.

\begin{IEEEproof}
	To prove Lemma \ref{thm1}, we first define $\tilde{\boldsymbol{H}}_{h}$ as
	\begin{equation}\label{PF3}
	\tilde{\boldsymbol{H}}_{h} \buildrel \Delta \over = \left [ {\boldsymbol{H}}_1,\ldots,{\boldsymbol{H}}_{h-1},{\boldsymbol{H}}_{h+1},\ldots,{\boldsymbol{H}}_H \right ]^{\textrm{T}},
	\end{equation}
	where $\boldsymbol{H}_{h-1}$ is the effective channels of all users in group $h-1$. We assume that the rank of $\tilde{\boldsymbol{H}}_{h}$ is $\tilde{L}_{k}$.
	Next, we introduce the use of BD method to represent the transmit beamforming matrices of the BS and the receive beamforming matrices of the users by the phase shifts of the IRS. To eliminate the inter-group interference, we define $\tilde{\boldsymbol{V}}_{h}^{\left(0\right)} \in \mathbb{C}^{N^\textrm{B} \times \left({N^\textrm{B}-\tilde{L}_{k}}\right)}$ as
	\begin{equation}\label{PF4}
	\tilde{\boldsymbol{V}}_{h}^{\left(0\right)}=\text{null} \left(\tilde{\boldsymbol{H}}_{h}\right),
	\end{equation}
	where $\text{null} \left(\tilde{\boldsymbol{H}}_{h}\right)$ represents that $\tilde{\boldsymbol{V}}_{h}^{\left(0\right)}$ lies in the $\text{null}$ space of $\tilde{\boldsymbol{H}}_{h}$.  Hence, we have $\tilde{\boldsymbol{H}}_{h} \tilde{\boldsymbol{V}}_{h}^{\left(0\right)} = \boldsymbol 0$. The self interference of each user can be eliminated by the SVD of ${\boldsymbol{H}}_{k} \tilde{\boldsymbol{V}}_{h}^{\left(0\right)}$, which is
	\begin{equation}\label{PF5}
	{\boldsymbol{H}}_{k} \tilde{\boldsymbol{V}}_{h}^{\left(0\right)}= \left [ {\boldsymbol{U}}_{k}^{\left(1\right)} , {\boldsymbol{U}}_{k}^{\left(2\right)} \right ] \left [ \begin{matrix}  {\boldsymbol{\Sigma}}_{k}^{\left(1\right)}&\boldsymbol{0}&\\\boldsymbol{0}&{\boldsymbol{\Sigma}}_{k}^{\left(2\right)}& \end{matrix} \!\!\!\!\! \right ] \left [ {\boldsymbol{V}}_{k}^{\left(1\right)} , {\boldsymbol{V}}_{k}^{\left(2\right)} \right ]^{\textrm{H}}.
	\end{equation}
	We assume that the rank of ${\boldsymbol{H}}_{k} \tilde{\boldsymbol{V}}_{h}^{\left(0\right)}$ is ${L}_{k}$, the column vectors of  $\boldsymbol{U}_{k}^{\left(1\right)} \in \mathbb{C}^{N^\textrm{U} \times \zeta}$, $\boldsymbol{U}_{k}^{\left(2\right)} \in \mathbb{C}^{N^\textrm{U} \times \left({{L}_{k}-\zeta}\right)}$, $\boldsymbol{V}_{k}^{\left(1\right)} \in \mathbb{C}^{\left({N^\textrm{B}-\tilde{L}_{k}}\right) \times \zeta}$, and $\boldsymbol{V}_{k}^{\left(2\right)} \in \mathbb{C}^{\left({N^\textrm{B}-\tilde{L}_{k}}\right) \times \left({{L}_{k}-\zeta}\right)}$ can form orthonormal sets, ${\boldsymbol{\Sigma}}_{k}^{\left(1\right)}\in  \mathbb{C}^{\zeta \times \zeta}$ and ${\boldsymbol{\Sigma}}_{k}^{\left(2\right)}\in  \mathbb{C}^{ \left({{L}_{k}-\zeta}\right) \times \left({{L}_{k}-\zeta}\right) }$ are diagonal matrices of singular values.
	${\boldsymbol{B}}_h\left({\boldsymbol{\nu}}\right)$ must be designed to cancel the inter-group interference as well as the interference of all users in this group. Thus, $\tilde{\boldsymbol{V}}_{h}^{\left(0\right)}$ and $\sum\limits_{i = 1}^K {\boldsymbol{V}_{i}^{\left(1\right)}}$ must be included in ${\boldsymbol{B}}_h\left({\boldsymbol{\nu}}\right)$, which can be given by
	\begin{equation}\label{PF6}
	{\boldsymbol{B}}_h\left({\boldsymbol{\nu}}\right) =\tilde{\boldsymbol{V}}_{h}^{\left(0\right)} \left(\frac{{\boldsymbol{V}}_1^{\left(1\right)}+\cdots+ {\boldsymbol{V}}_K^{\left(1\right)} }{\sqrt{\left | \mathcal H_h \right |}}\right) {{\boldsymbol{G}}_{h}^{1/2}},
	\end{equation}
	where $ \left | \mathcal H_h \right | $ is the number of users in group $h$, $\frac{1}{\sqrt{\left | \mathcal H_h \right |}}$ ensures that the power of $\left(\frac{{\boldsymbol{V}}_1^{\left(1\right)}+\cdots+ {\boldsymbol{V}}_K^{\left(1\right)} }{\sqrt{\left | \mathcal H_h \right |}}\right)$ is unit, and ${{\boldsymbol{G}}_{h}}$ is the power allocation matrix in group $h$.
 To eliminate the interference of all users in this group, the fully digital receive beamforming matrix ${\boldsymbol{J}}_{k}$ of user $k$ in group $h$ is written as    	
	\begin{equation}\label{PF7+}
	\begin{split}
	{\boldsymbol{J}}_{k}\left({\boldsymbol{\nu}}\right)&={\boldsymbol{U}}_k^{\left(1\right)}.
	\end{split}
	\end{equation}
    Substituting ${{\boldsymbol{G}}_{h}}$ into (\ref{PF6}), we have
	\begin{equation}\label{PF8}
	{\boldsymbol{B}}_h\left({\boldsymbol{\nu}}\right) = \tilde{\boldsymbol{V}}_{h}^{\left(0\right)} \left(\frac{{\boldsymbol{V}}_1^{\left(1\right)}+\cdots+ {\boldsymbol{V}}_K^{\left(1\right)} }{\sqrt{\left | \mathcal H_h \right |}}\right) {\sqrt{\frac{P}{{H\zeta}}}}.
	\end{equation}
	This completes the proof.
\end{IEEEproof}

From Lemma \ref{thm1}, we can see that ${\boldsymbol{B}}_h\left({\boldsymbol{\nu}}\right)$ mainly depends on the orthogonal bases of the null space of users in other groups, the orthogonal bases of the subspace of users in group $h$, the maximum transmit power of the BS and number of groups, ${\boldsymbol{J}}_{k}\left({\boldsymbol{\nu}}\right)$ depends on the effective  channel of user $k$ and the orthogonal bases of the null space of users in other groups.

\subsubsection{Simplification of Problem (\ref{PF2})}
Based on the Lemma~\ref{thm1}, the interference caused by other groups ${J_{ik,h}}$ and user itself ${I_{ik,h}}$
can be eliminated by the fully digital transmit beamforming matrix ${\boldsymbol{B}}_h\left({\boldsymbol{\nu}}\right)$ and receive beamforming matrix ${\boldsymbol{J}}_{k}\left({\boldsymbol{\nu}}\right)$. Substituting ${\boldsymbol{B}}_h\left({\boldsymbol{\nu}}\right)$ and  ${\boldsymbol{J}}_{k}\left({\boldsymbol{\nu}}\right)$ into (\ref{PF2}), the achievable data rate of user $k$ in group $h$ can be rewritten as follows:
\begin{equation}\label{PF9+}
\begin{split}
&{R_{k,h}}\left({{\boldsymbol{B}}_h\left({\boldsymbol{\nu}}\right),  {\boldsymbol{J}}_{k}\left({\boldsymbol{\nu}}\right),  {\boldsymbol{\nu}}}\right)= \\
&\;\;\;\;\;\;W\sum_{i=1}^{\zeta} \log_2 \left ( 1 + {\left |  \left(\boldsymbol{j}_{k,i}\left({\boldsymbol{\nu}}\right)\right)^{\textrm{H}} \boldsymbol{H}_{k}  \bar{\boldsymbol{b}}_{h,i}\left({\boldsymbol{\nu}}\right) \right |^2}    /    {\sigma^2} \right ),\\
\end{split}
\end{equation}
Let $\frac{\left |  \left(\boldsymbol{j}_{k,i}\left({\boldsymbol{\nu}}\right)\right)^{\textrm{H}}  \boldsymbol{H}_{k}  \bar{\boldsymbol{b}}_{h,i}\left({\boldsymbol{\nu}}\right) \right |^2}        {\sigma^2}={\lambda _i}$, (\ref{PF9+}) can be rewritten by
\begin{equation}\label{PF9++}
\begin{split}
&{R_{k,h}}\left({{\boldsymbol{B}}_h\left({\boldsymbol{\nu}}\right),  {\boldsymbol{J}}_{k}\left({\boldsymbol{\nu}}\right),  {\boldsymbol{\nu}} }\right) \\  
&=W\sum_{i=1}^{\zeta} \log_2 \left ( 1 + {\lambda _i} \right ),\\
&=W\log_2 \left ( 1 + {\lambda _1} \right )+\ldots+\log_2 \left ( 1 + {\lambda _{\zeta}} \right ),\\
&=W\log_2 \left(\left ( 1 + {\lambda _1} \right )*\ldots*\left ( 1 + {\lambda _{\zeta}} \right )\right),\\
&=W\log_2 \left |  \begin{matrix} 1 + {\lambda _1}&\cdots&0&\\\vdots&\ddots&\vdots&\\0&\cdots&1 + {\lambda _{\zeta}} \end{matrix} \!\!\!\!\!  \right|.
\end{split}
\end{equation}
Therefore, we have
\begin{equation}\label{PF9+++}
\begin{split}
&{R_{k,h}}\left({{\boldsymbol{B}}_h\left({\boldsymbol{\nu}}\right),  {\boldsymbol{J}}_{k}\left({\boldsymbol{\nu}}\right),  {\boldsymbol{\nu}} }\right) \\
&=W\log_2 \text{det} \left ( \boldsymbol{I}_{\zeta} + {\left |  \left({\boldsymbol{J}}_{k}\left({\boldsymbol{\nu}}\right)\right)^{\textrm{H}} \boldsymbol{H}_{k}  \boldsymbol{B}_h\left({\boldsymbol{\nu}}\right) \right |^2}     /   {\sigma^2} \right ),
\end{split}
\end{equation}
where $\text{det}(\cdot)$  represents the determinant of a square matrix, and $\boldsymbol{I}_{\zeta}$ is an $\zeta \times \zeta$ identity matrix.
Substituting ${\boldsymbol{B}}_h\left({\boldsymbol{\nu}}\right)$ and  ${\boldsymbol{J}}_{k}\left({\boldsymbol{\nu}}\right)$ into (\ref{PF9+++}), we have
\begin{equation}\label{PF9++++}
\begin{split}
&{R_{k,h}}\left({{\boldsymbol{B}}_h\left({\boldsymbol{\nu}}\right),  {\boldsymbol{J}}_{k}\left({\boldsymbol{\nu}}\right),  {\boldsymbol{\nu}} }\right) \\
&= W\log_2 \text{det}\! \left (\! \boldsymbol{I}_{\zeta} \!+\! \frac{\left |  \!\left(\!{\boldsymbol{U}}_k^{\left(1\right)}\right)^{\textrm{H}} \!\!\boldsymbol{H}_{k}  \!\tilde{\boldsymbol{V}}_{h}^{\left(0\right)} \!\!\left(\!\frac{\sum\limits_{i = 1}^K {\boldsymbol{V}_{i}^{\left(1\right)}} }{\sqrt{\left | \mathcal H_h \right |}}\!\right)\! {\sqrt{\frac{P}{{H\zeta}}}} \right |^2}        {\sigma^2} \right ).
\end{split}
\end{equation}
Since ${\boldsymbol{H}}_i \left({\boldsymbol{H}}_k\right)^{\textrm{H}}=\boldsymbol 0 $ $\left(i \neq k\right)$, we have ${\boldsymbol{H}}_k \left({\boldsymbol{V}}_i^{\left(1\right)}\right)^{\textrm{H}}=\boldsymbol 0$ $\left(i \neq k\right)$.
Substituting (\ref{PF5}) into (\ref{PF9++++}), we have
\begin{equation}\label{PF9+++++}
\begin{split}
&{R_{k,h}}\left({{\boldsymbol{B}}_h\left({\boldsymbol{\nu}}\right),  {\boldsymbol{J}}_{k}\left({\boldsymbol{\nu}}\right),  {\boldsymbol{\nu}} }\right) \\  
&= W\log_2 \text{det} \left ( \boldsymbol{I}_{\zeta} + {\frac{P}{{\left | \mathcal H_h \right |}H\zeta\sigma^2 }} \left({\boldsymbol{\Sigma}}_{k}^{\left(1\right)}\right)^2  \right ),
\end{split}
\end{equation}
Then, the optimization problem in (\ref{PF2}) can be transformed as
\begin{equation}\label{PF10}
\begin{split}
\mathop {\max }\limits_{ {\boldsymbol{\nu}} }  &   \sum\limits_{h = 1}^H {\mathop {\min }\limits_{k \in {\mathcal H_h}} \left\{ W {\log_2 \text{det} \left ( \boldsymbol{I}_{\zeta} + {\frac{P}{{\left | \mathcal H_h \right |}H\zeta\sigma^2 }} \left({\boldsymbol{\Sigma}}_{k}^{\left(1\right)}\right)^2  \right )}\right\} } \\
\textrm{s.t.} &\quad \left(\ref{PF1}\textrm{c}\right).
\end{split}
\end{equation}

\subsection{Phase Optimization with Manifold Method}\label{Passive Beamforming Design with Manifold}

\subsubsection{Approximation of ${\boldsymbol{\Sigma}}_{k}^{\left(1\right)}$}
Since ${\boldsymbol{\Sigma}}_{k}^{\left(1\right)}$ in (\ref{PF10}) is unknown, we use the function of phase shift to represent ${\boldsymbol{\Sigma}}_{k}^{\left(1\right)}$, which is proved in Theorem \ref{thm1}.
\newtheorem{thm}{\bf Theorem}
\begin{thm}\label{thm1}
 ${\boldsymbol{\Sigma}}_{k}^{\left(1\right)}\left(i,j\right) \approx {\beta _i} {\alpha _j}{\boldsymbol{\nu}}^{\textrm{H}} {{\boldsymbol{c}}^{ij}}$, where ${{\boldsymbol{c}}^{ij}}= \left(\boldsymbol{a}\left({\theta}_{i}^{{\textrm{D}}},{\eta}_{i}^{{\textrm{D}}}\right)\right)^{{\textrm{*}}}    \circ    \boldsymbol{a}\left({\theta}_{j}^{{\textrm{A}}},{\eta}_{j}^{{\textrm{A}}}\right) $ with $\left(\boldsymbol{a}\left({\theta}_{i}^{{\textrm{D}}},{\eta}_{i}^{{\textrm{D}}}\right)\right)^{{\textrm{*}}}$ being the conjugate of $\left(\boldsymbol{a}\left({\theta}_{i}^{{\textrm{D}}},{\eta}_{i}^{{\textrm{D}}}\right)\right)$ and $\circ$ being the Hadamard product.
\end{thm}

\begin{IEEEproof}
	See Appendix  \ref{Proof of thm1}.
\end{IEEEproof}

From Theorem \ref{thm1}, we can see that ${\boldsymbol{\Sigma}}_{k}^{\left(1\right)}$ depends on the distance ${\alpha _j}$ between the BS and the IRS, the distance ${\beta _i}$ between the IRS and user $k$, the angle $\boldsymbol{a}\left({\theta}_{j}^{{\textrm{A}}},{\eta}_{j}^{{\textrm{A}}}\right)$ from the BS to the IRS, the phase shifts of the IRS, and the angle $\boldsymbol{a}\left({\theta}_{i}^{{\textrm{D}}},{\eta}_{i}^{{\textrm{D}}}\right)$ from the IRS to user $k$.

\subsubsection{Problem Transformation}
Based on Theorem \ref{thm1}, the optimization problem (\ref{PF10}) can be rewritten as
\begin{subequations}\label{PF14}  \vspace{-.3em}
	\begin{align}
	\mathop {\max }\limits_{ \boldsymbol{\nu} }    \sum\limits_{h = 1}^H & {\mathop {\min }\limits_{k \in {\mathcal H_h}}   W \left\{\sum\limits_{i = 1}^{\zeta} {\log_2  \left ( 1 + {\frac{P}{{\left | \mathcal H_h \right |}H\zeta\sigma^2 }}  \left | \boldsymbol{D}_{k}\left(i,i\right) \right |^2  \right )}\right\} } \tag{\ref{PF14}}\\
	\textrm{s.t.} \quad \:
	&\! \! \left(\ref{PF1}\textrm{c}\right),\notag\\
	&{\left| {{d_{ij}}} \right|} = \left| {{\boldsymbol{\nu}}^H{{\boldsymbol{c}}^{ij}}} \right|  \textless \tau,  \quad \forall i \ne j,
	\end{align}
\end{subequations}
where ${d_{ii}} = {\boldsymbol{\nu}}^H{{\boldsymbol{c}}^{ii}}$, $\boldsymbol{D}_{k}\left(i,i\right)={\alpha _i} {\beta _i} {d_{ii}}$ ($i \in \left\{ 1, \ldots ,{\zeta}\right\}$), and $\tau$ is a small positive value. Constraint (\ref{PF14}a)  is to make sure that $\boldsymbol{D}_{k}$ is approximately a non-square diagonal matrix such that ${\boldsymbol{H}}_{k} \tilde{\boldsymbol{V}}_{h}^{\left(0\right)}=\boldsymbol{A}_{k} \boldsymbol{D}_{k} \left(\boldsymbol{A}\right)^{{\textrm{H}}} \left [\boldsymbol{z}_{\left(K-{\left | \mathcal H_h \right |}\right){\zeta}+1}; \ldots; \boldsymbol{z}_{K{\zeta}}  \right ]$ can be treated as an approximation of the truncated SVD of  ${\boldsymbol{H}}_{k} \tilde{\boldsymbol{V}}_{h}^{\left(0\right)}$, where $\boldsymbol{z}_{\left(K-{\left | \mathcal H_h \right |}\right){\zeta}+1}$ denotes row ${\left(K-{\left | \mathcal H_h \right |}\right){\zeta}+1}$ of matrix ${\boldsymbol{Z}}$.  Constraint (\ref{PF14}a) can be removed and this omission does not affect the validity of our proposed solution\cite{9234098}.  Hence, problem (\ref{PF14}) can be rewritten as follows:

\begin{equation}\label{PF15}
\begin{split}
\mathop {\max }\limits_{ \boldsymbol{\nu} }  &   \sum\limits_{h = 1}^H {\mathop {\min }\limits_{k \in {\mathcal H_h}} W \left\{  \sum\limits_{i = 1}^{\zeta} {\log_2  \left ( 1 + {\frac{P}{{\left | \mathcal H_h \right |}H\zeta\sigma^2 }}  \left | \boldsymbol{D}_{k}\left(i,i\right) \right |^2  \right )}\right\} }\\
\textrm{s.t.}& \quad \; \left(\ref{PF1}\textrm{c}\right).
\end{split}
\end{equation}
Substituting $\boldsymbol{D}_{k}\left(i,i\right)={\alpha _i} {\beta _i} {\boldsymbol{\nu}}^H{{\boldsymbol{c}}^{ii}}$ into (\ref{PF15}), the problem (\ref{PF15}) can be transformed as follows:
\begin{equation}\label{PF17} \vspace{-.3em}
\begin{split}
\mathop {\max }\limits_{ \boldsymbol{\nu}}  & \quad  \sum\limits_{h = 1}^H {\mathop {\min }\limits_{k \in {\mathcal H_h}} W \left\{  \sum\limits_{i = 1}^{\zeta} {\log_2  \left ( 1 +{b _i} {\boldsymbol{\nu}}^{\textrm{H}} {{\boldsymbol{C}}^{ii}} {\boldsymbol{\nu}} \right)}\right\} } \\
\textrm{s.t.} &\quad \; \left(\ref{PF1}\textrm{c}\right),
\end{split}
\end{equation}
where ${{\boldsymbol{C}}^{ii}} \buildrel \Delta \over ={{\boldsymbol{c}}^{ii}} \left({\boldsymbol{c}}^{ii} \right)^{\textrm{H}}$ and $ {b _i} \buildrel \Delta \over =  {\frac{P}{{\left | \mathcal H_h \right |}H\zeta\sigma^2 }}  \left | {\alpha _i} {\beta _i}  \right |^2$.

\subsubsection{Solution of Problem (\ref{PF17}) }

Since constraint (\ref{PF1}c) has a manifold structure, problem (\ref{PF17}) can be regarded as a manifold-constrained optimization problem. Next, we introduce the use of a manifold method \cite{Optimization2009} to solve problem (\ref{PF17}). In particular, we first introduce the definition of a tangent space. Then, similar to the gradient in Euclidean space, we introduce the gradient on the manifold, called the Riemannian gradient.  Finally, problem (\ref{PF17}) is solved by an iterative method using the Riemannian gradient. 

To solve problem (\ref{PF17}), we first rewrite the objective function as
\begin{equation}\label{PF18}
\begin{split}
f \left(\boldsymbol{\nu}\right) \buildrel \Delta \over =-\sum\limits_{h = 1}^H {\mathop {\min }\limits_{k \in {\mathcal H_h}} W\left\{  \sum\limits_{i = 1}^{\zeta}  {\log_2  \left ( 1 +{b _i} {\boldsymbol{\nu}}^{\textrm{H}} {{\boldsymbol{C}}^{ii}} {\boldsymbol{\nu}} \right)} \right\} }.
\end{split}
\end{equation}
Let ${{\boldsymbol{\nu}}_{n}}$  be the value at iteration $n$. Based on (\ref{PF18}), the Euclidean gradient of the objective function $f \left(\boldsymbol{\nu}\right) $   at point ${{\boldsymbol{\nu}}_{n}}$ is given by 
\begin{equation}\label{PF071810}
\begin{split}
\nabla f\left({{\boldsymbol{\nu}}_n}\right) & =-\sum\limits_{h = 1}^H {\mathop {\min }\limits_{k \in {\mathcal H_h}} \left\{ \nabla \left( {R_{k,h}}\left({{\boldsymbol{B}}_h\left({\boldsymbol{\nu}_n}\right),  {\boldsymbol{J}}_{k}\left({\boldsymbol{\nu}_n}\right),  {\boldsymbol{\nu}_n} }\right) \right)\right\} },\\
& =-\sum\limits_{h = 1}^H {\mathop {\min }\limits_{k \in {\mathcal H_h}}W \left\{ \sum\limits_{i = 1}^{\zeta}  \frac{1}{{\textrm {ln2} }} \frac{2{b _i} {{\boldsymbol{C}}^{ii}} {{\boldsymbol{\nu}}_{n}}   }  {1+{b _i}  \left ( {\boldsymbol{\nu}}_{n}\right)^{\textrm{H}} {{\boldsymbol{C}}^{ii}} {{\boldsymbol{\nu}}_{n}}}  \right\} }.
\end{split}
\end{equation}

To introduce the Riemannian gradient, we first define a tangent space of of an oblique manifold $\mathcal Q$ at point ${\boldsymbol{\nu}}_{n}$ as 
\begin{equation}\label{PF07189}
\begin{split}
{{{ T}_{{\boldsymbol{\nu}}_{n}}}\mathcal Q}=\left\{{\boldsymbol{u} \in \mathbb{C}^{G}} | \left[\boldsymbol{u} {{\boldsymbol{\nu}}_{n}^{\textrm{H}}}         \right]_{g,g} = 0,  \forall g \in \mathcal G = \left\{ 1,2,...,G      \right\}      \right\}.
\end{split}
\end{equation}
From (\ref{PF07189}), we see that the tangent space containts all the  tangent vectors of $\mathcal Q$ at ${\boldsymbol{\nu}}_{n}$.

 The Riemannian gradient of the objective function $f \left(\boldsymbol{\nu}\right) $  at point ${{\boldsymbol{\nu}}_{n}}$ can be obtained by orthogonally projecting the Euclidean gradient $\nabla f\left({{\boldsymbol{\nu}}_n}\right)$ onto the tangent space ${{{ T}_{{\boldsymbol{\nu}}_{n}}}\mathcal Q}$, which is  given by
\begin{equation}\label{PF07181035}
\begin{split}
{\mathcal G _{{\boldsymbol{\nu}}_{n}} \mathcal Q}
& =\nabla f\left({{\boldsymbol{\nu}}_n}\right)-  {\textrm {Real}}\left\{ \nabla f\left({{\boldsymbol{\nu}}_n}\right) \circ {\left({\boldsymbol{\nu}}_{n}^{\textrm{T}}\right)^ {\textrm{H}} }   \right\}  \circ {\boldsymbol{\nu}}_{n},
\end{split}
\end{equation}
where ${\textrm {Real}}(\boldsymbol{M})$ is the real part of $\boldsymbol{M}$ and  ${\textrm {Real}}\left\{ \nabla f\left({{\boldsymbol{\nu}}_n}\right) \circ {\left({\boldsymbol{\nu}}_{n}^{\textrm{T}}\right)^ {\textrm{H}} }   \right\}  \circ {\boldsymbol{\nu}}_{n}$      is the projected gradient of Euclidean gradient $\nabla f\left({{\boldsymbol{\nu}}_n}\right)$ on the tangent space ${{{ T}_{{\boldsymbol{\nu}}_{n}}}\mathcal Q}$.

Given the Riemannian gradient, we use the optimization method in Euclidean space to solve the manifold-constrained optimization problem\cite{9024490}. The update of  ${\boldsymbol{\nu}}_{n}$ is
\begin{equation}\label{PF21}
\begin{split}
\bar{\boldsymbol{\nu}}_{n}={\boldsymbol{\nu}}_{n}-{\tilde \lambda _n} {\mathcal G  _{{\boldsymbol{\nu}}_{n}} \mathcal Q}, 
\end{split}
\end{equation}
where ${\tilde \lambda _n}$ is the step size. To ensure that the updated value of ${\boldsymbol{\nu}_n}$  lies in the feasible set, we have 
\begin{equation}\label{PF22}
\begin{split}
{\boldsymbol{\nu}}_{n+1}=\bar{\boldsymbol{\nu}}_{n} \circ \frac{1}{\left | {\bar{\boldsymbol{\nu}}_{n} }\right | }.
\end{split}
\end{equation}
The detailed process of using the manifold-based method to solve problem (\ref{PF17}) is given in Algorithm \ref{071811}. Given ${\boldsymbol{\nu}}$, we can use Lemma \ref{thm1} to calculate ${\boldsymbol{B}}_h\left({\boldsymbol{\nu}}\right)$  and ${\boldsymbol{J}}_{k}\left({\boldsymbol{\nu}}\right)$.

\begin{algorithm}[t]
	\caption{Proposed Design for Solving (\ref{PF17})}
	\begin{algorithmic}[1]\label{071811}
		\STATE Initialize ${\boldsymbol{\nu}}_{0} \in {\mathcal Q}$.
		\STATE Obtain ${d_{ij}}$ according to  (\ref{PF11+}). 
		\STATE Obtain ${{\boldsymbol{C}}^{ii}}$ and ${b _i}$ according to (\ref{PF17}). 
		\REPEAT
		\STATE Compute the Euclidean gradient using (\ref{PF071810}).
		\STATE Compute the Riemannian gradient by (\ref{PF07181035}).
		\STATE Update ${\boldsymbol{\nu}}_{n+1}$ by (\ref{PF22}).
		\UNTIL the objective function converges.
		\STATE ${\boldsymbol{\Phi}}=\text{diag}({\boldsymbol{\nu} }^{\textrm{H}})$.
	\end{algorithmic}
\end{algorithm}

\subsection{Optimization of the Transmit Beamforming Matrices of the BS and the Receive Beamforming Matrices of  Users}\label{the transmit beamforming matrices of the BS, the receive beamforming matrices of the users}
Given ${\boldsymbol{B}}_h\left({\boldsymbol{\nu}}\right)$  and ${\boldsymbol{J}}_{k}\left({\boldsymbol{\nu}}\right)$,  we next introduce the use of the algorithm in \cite{8646553} to optimize ${\boldsymbol{F}^\textrm{B}}$, ${\boldsymbol{F}^\textrm{R}}$, $\boldsymbol{W}_{k}^{\textrm{B}}$, and $\boldsymbol{W}_{k}^{\textrm{R}}$. 

   Since we have assumed that $\boldsymbol{B} = \left[\boldsymbol{B}_1,\ldots,\boldsymbol{B}_H \right] \in \mathbb{C}^{N^\textrm{B} \times H\zeta}$ is a fully digital transmit beamforming matrix which has the same size as the hybrid transmit beamforming matrix $\boldsymbol{F}^\textrm{R}\boldsymbol{F}^\textrm{B}$ and we have also replaced $\boldsymbol{F}^\textrm{R}\boldsymbol{F}^\textrm{B}$ with $\boldsymbol{B}$ in problem (\ref{PF2}),
 the problem of optimizing ${\boldsymbol{F}^\textrm{B}}$ and ${\boldsymbol{F}^\textrm{R}}$ can be formulated as 
\begin{subequations}   \label{PF0720}
	\begin{align}
	&\mathop {\min}_{\boldsymbol{F}^\textrm{R},\boldsymbol{F}^\textrm{B} }     \left\|\boldsymbol{B}-\boldsymbol{F}^\textrm{R}\boldsymbol{F}^\textrm{B}\right\|_F^2 \tag{\theequation} \\
	&\quad\: \ \ {\rm{      }}\left| \boldsymbol{F}^\textrm{R}\left(i,j\right) \right|  = 1, \forall i,j,
	\end{align}
\end{subequations}

Due to the unit modulus constraints of (\ref{PF0720}a),  vector $\boldsymbol{x} = \textrm{v}\left(\boldsymbol{F}^\textrm{R}\right)$ forms a complex circle manifold $\left\{  \boldsymbol{x}  \in \mathbb{C}^w : {\rm{      }}\left| \boldsymbol{x}_1 \right| = \left| \boldsymbol{x}_2   \right|  = \cdots = \left| \boldsymbol{x}_w   \right| = 1 \right\}$, where $\textrm{v}\left(\boldsymbol{F}^\textrm{R}\right)$  is the vectorization of $\boldsymbol{F}^\textrm{R}$ and $w = N^\textrm{B} M^\textrm{B}$.  Hence, problem (\ref{PF0720}) can be transformed to  an unconstrained optimization problem on manifolds. The  iterative algorithm used to solve problem (\ref{PF17}) can be used to solve it. In particular,  $\boldsymbol{x}^{n}$ can be updated using  (\ref{PF21}) and (\ref{PF22}).
Given the updated $\boldsymbol{x}$ with $\boldsymbol{x} = \textrm{v}\left(\boldsymbol{F}^\textrm{R}\right)$, the update of the RF transmit beamforming matrix at  iteration $n$ can be expressed as   
\begin{subequations}   \label{PF07201845}
	\begin{align}
\boldsymbol{F}^\textrm{R}_n = \textrm{v}^{-1}\left(\boldsymbol{x}^n\right), \tag{\theequation}
	\end{align}
\end{subequations}
where $\textrm{v}^{-1}\left(\boldsymbol{x}^n\right)$ is the inverse-vectorization of $\boldsymbol{x}^n$.
Given $\boldsymbol{F}^\textrm{R}_n$, the optimization problem in (\ref{PF0720}) can be simplified as follows:
\begin{equation}   \label{PF0724}
	\mathop {\min}_{\boldsymbol{F}^\textrm{B} }     \left\|\boldsymbol{B}-\boldsymbol{F}^\textrm{R}_n\boldsymbol{F}^\textrm{B}\right\|_F^2. 
\end{equation}
Since problem (\ref{PF0724})  is a least-square optimization problem,  $\boldsymbol{F}^\textrm{B}$ at iteration $n$ can be given by
\begin{subequations}   \label{PF07201851}
	\begin{align}
	\boldsymbol{F}^\textrm{B}_n = \left(\boldsymbol{F}^\textrm{R}_n \right)^{\dagger}  \boldsymbol{B},\tag{\theequation}
	\end{align}
\end{subequations}
where $\left(\boldsymbol{F}^\textrm{R}_n \right)^{\dagger} $ is the Moore-Penrose pseudo inverse of $\boldsymbol{F}^\textrm{R}_n$. At convergence, ${\boldsymbol{F}^\textrm{R}}$ is expressed as
\begin{subequations}   \label{PF07201903}
	\begin{align}
	{\boldsymbol{F}^\textrm{R}} = \textrm{v}^{-1}\left(\boldsymbol{x}^{n+1}\right). \tag{\theequation}
	\end{align}
\end{subequations}          
To satisfy constraint    (\ref{PF1}a), ${\boldsymbol{F}^\textrm{B}}$ is expressed as
\begin{subequations}   \label{PF07201905}
	\begin{align}
	{\boldsymbol{F}^\textrm{B}} =\frac {\sqrt{P}} {\left\|\boldsymbol{F}^\textrm{R}\boldsymbol{F}^\textrm{B}_n\right\|_F}  \boldsymbol{F}^\textrm{B}_n.
	 \tag{\theequation}
	\end{align}
\end{subequations}

Similarly, given ${\boldsymbol{J}}_{k}\left({\boldsymbol{\nu}}\right)$, the problem of optimizing $\boldsymbol{W}_{k}^{\textrm{B}}$ and $\boldsymbol{W}_{k}^{\textrm{R}}$ can be given as follows:
\begin{subequations}   \label{PF0728}
	\begin{align}
	&\mathop {\min}_{\boldsymbol{W}_{k}^{\textrm{R}}, \boldsymbol{W}_{k}^{\textrm{B}} }     \left\|\boldsymbol{J}_{k}-\boldsymbol{W}_{k}^{\textrm{R}} \boldsymbol{W}_{k}^{\textrm{B}}\right\|_F^2 \tag{\theequation} \\
	&\quad\: \ \ \ {\rm{      }}\left| \boldsymbol{W}_{k}^{\textrm{R}}\left(i,j\right) \right| = 1, \forall i,j.
	\end{align}
\end{subequations}
Since this optimization problem is similar to the problem in (\ref{PF0720}), we can use the same method used to solve problem (\ref{PF0720}) to optimize $\boldsymbol{W}_{k}^{\textrm{B}}$ and $\boldsymbol{W}_{k}^{\textrm{R}}$.

\subsection{Complexity Analysis}
	
	The proposed algorithm for solving problem (\ref{PF1}) is summarized in Algorithm \ref{PAP}. The complexity of Algorithm \ref{PAP} lies in the calculation of (\ref{PF6}), solving problem (\ref{PF17}), and using the algorithm in \cite{8646553} to find $\hat{\boldsymbol{F}^\textrm{B}}, \hat{\boldsymbol{F}^\textrm{R}}$, $ \hat{\boldsymbol{W}_{k}^{\textrm{B}}}$, and $\hat{\boldsymbol{W}_{k}^{\textrm{R}}} $.
	
	The complexity of calculating (\ref{PF6}) is $\mathcal O\left( \left(N^\textrm{B}\right) ^{3} + {\left | \mathcal H_h \right |}\left(N^\textrm{B}\right) ^{2}\zeta\right)$. The complexity  of solving problem (\ref{PF17}) lies in computing the Euclidean gradient of the objective function in (\ref{PF17}) at each iteration, which involves the complexity of $\mathcal O\left( H {M}^{2} \zeta \right)$. Hence, the total complexity of solving problem (\ref{PF17}) is $\mathcal O\left( H {M}^{2} \zeta {S_1}\right)$, where ${S_1}$ is the number of iterations of using the manifold method to solve problem (\ref{PF17}). The complexity of using the algorithm in \cite{8646553} to find $\hat{\boldsymbol{F}^\textrm{B}}, \hat{\boldsymbol{F}^\textrm{R}}$,  $\hat{\boldsymbol{W}_{k}^{\textrm{B}}}$, and $\hat{\boldsymbol{W}_{k}^{\textrm{R}}}$ is $\mathcal O\left( N^\textrm{B} M^\textrm{B} \zeta {S_2}+N^\textrm{U} M^\textrm{U} \zeta {S_2} \right)$, where ${S_2}$ is the number of iterations required to converge. Hence, the total complexity of solving problem (\ref{PF1}) is $ \mathcal O\left(\left(N^\textrm{B}\right) ^{3} + {\left | \mathcal H_h \right |}\left(N^\textrm{B}\right) ^{2}\zeta + H {M}^{2} \zeta {S_1}+N^\textrm{B} M^\textrm{B} \zeta {S_2}+N^\textrm{U} M^\textrm{U} \zeta {S_2}   \right) \\ \approx O\left(\left(N^\textrm{B}\right) ^{3} + H {M}^{2} \zeta {S_1} \right)$.

\begin{algorithm}[t]
	\begin{small}
	\caption{Proposed Scheme  for Solving Problem (\ref{PF1})}
	\begin{algorithmic}[1]\label{PAP}
     	\STATE {\textbf{Input:} } $\boldsymbol{H}^\textrm{B}, \boldsymbol{H}^\textrm{R}_{k}, \zeta, P, \sigma^2$.
		\STATE Calculate  ${\boldsymbol{B}}_h\left({\boldsymbol{\nu}}\right)$ and  ${\boldsymbol{J}}_{k}\left({\boldsymbol{\nu}}\right)$ by Lemma \ref{thm1}.
		\STATE Find the phase shift $\hat{\boldsymbol{\nu}}$ of the IRS by solving  problem (\ref{PF17}).
		\STATE Obtain ${\boldsymbol{B}}_h\left(\hat{\boldsymbol{\nu}}\right)$ and  ${\boldsymbol{J}}_{k}\left(\hat{\boldsymbol{\nu}}\right)$ by Lemma \ref{thm1}.
		\STATE Calculate $\hat{\boldsymbol{F}^\textrm{B}}, \hat{\boldsymbol{F}^\textrm{R}},$  $ \hat{\boldsymbol{W}_{k}^{\textrm{B}}}$, and $\hat{\boldsymbol{W}_{k}^{\textrm{R}}} $ by the algorithm in \cite{8646553}.
		\STATE {\textbf{Output:} } $\hat{\boldsymbol{\nu}}, \hat{\boldsymbol{F}^\textrm{B}}, \hat{\boldsymbol{F}^\textrm{R}}$,   $\hat{\boldsymbol{W}_{k}^{\textrm{B}}}$,  $\hat{\boldsymbol{W}_{k}^{\textrm{R}}}$.
	\end{algorithmic}
\end{small}
\end{algorithm}

\subsection{Convergence Analysis}
\begin{thm} \label{thm2}
	Assume that there exists $ \boldsymbol{\nu}^*$ such that $\mathcal  G_{\boldsymbol{\nu}^*} \mathcal  Q=\boldsymbol 0$. 
	Then, there exists a neighborhood $\mathcal U$ of  $ \boldsymbol{\nu}^*$  in $ \mathbb{C}^G$, such that for all  $\boldsymbol{\nu}_0\in\mathcal U$, Algorithm 1 generates an infinite sequence $\{\boldsymbol{\nu}_n\}$ converging to $ \boldsymbol{\nu}^*$.
\end{thm}

\begin{IEEEproof}
According to \cite[Theorem 6.3.2]{Optimization2009}, there exists $\gamma_R$ such that
\begin{equation}
\|\boldsymbol{\nu}_{n+1}-\bar {\boldsymbol{\nu}}_n\|
\leq \gamma_R
\|{\boldsymbol{\nu}}_n -{\boldsymbol{\nu}}^*\|^2.
\end{equation} 
Based on Algorithm 1, we further have
\begin{align}
\|\boldsymbol{\nu}_{n+1}- {\boldsymbol{\nu}}^*\| 
&\leq \|\boldsymbol{\nu}_{n+1}-\bar {\boldsymbol{\nu}}_n\|
+\|\bar {\boldsymbol{\nu}}_n-{\boldsymbol{\nu}}^*\|
 \nonumber\\
 &
 \leq  \|  {\boldsymbol{\nu}}_n-{\boldsymbol{\nu}}^*
 -\tilde{\lambda}_n\mathcal  G_{\boldsymbol{\nu}^*} \mathcal  Q
 \|
 \nonumber\\
  &+\gamma_R
 \|{\boldsymbol{\nu}}_n -{\boldsymbol{\nu}}^*\|^2
 \nonumber\\
&=\gamma_T
\|{\boldsymbol{\nu}}_n -{\boldsymbol{\nu}}^*\|^2 
+\gamma_R
\|{\boldsymbol{\nu}}_n -{\boldsymbol{\nu}}^*\|^2 
\end{align}
where the last inequality follows from the 
Lipschitz-continuous differential of function $f(\boldsymbol \nu)$ and $\gamma_T>0$ is a parameter related to the step-size $\tilde{\lambda}_n$ \cite[Theorem 6.3.2] {Optimization2009}.
\end{IEEEproof}

\section{Simulation Results}\label{SIMULATION RESULTS}
 In our simulations, the coordinates of the BS and the IRS are  (2m, 0m, 10m) and (0m, 148m, 10m), respectively. All users are randomly distributed in a circle centered at (7m, 148m, 1.8m) with a radius being 10 m. The bandwidth is set to 251.1886 MHz. The values of other parameters are defined in Table II. For comparison purposes, we consider  five baselines: a) the fully digital beamforming matrices of the BS and the users as well as the phase of each element of the IRS are optimized by the proposed scheme, which is the  solution for problem (\ref{PF2}),  b) the hybrid beamforming matrices of the BS and the users are determined by the proposed algorithm and the  phase shifts of the IRS are randomly determined, c) the fully digital beamforming matrices of the BS and the users are determined by the proposed scheme while the  phase shifts of the IRS are randomly determined, d) the hybrid beamforming matrices of the BS and the users are determined by the  algorithm in \cite{9234098} and the phase of each element of the IRS are optimized by the proposed scheme, e) the fully digital beamforming matrices of the BS and the users are determined by the  scheme in \cite{9234098} as well as the phase of each element of the IRS are optimized by the proposed scheme.
\begin{table}[t]
	\renewcommand{\arraystretch}{1.4}
	\centering
	\caption{Simulation Parameters}
	\label{table2}
	\begin{tabular}{|c|c|c|c|}
		\hline
		\textbf{Parameters}                    & \textbf{Values} &\textbf{Parameters}                    & \textbf{Values}     \\ \hline
		$M$  & $16\times 16$ &  $N^\textrm{B}$          & 64  \\ \hline
		$N^\textrm{U}$           & 64          &$M^\textrm{B}$            & 8              \\ \hline
		$M^\textrm{U}$                           & 4      &${\zeta}$  & 4            \\ \hline
		$Y=L$& 7           &${G_\textrm{t}}$ & 24.5 dBi            \\ \hline
		${G_\textrm{r}}$                    & 0 dBi    &$\sigma^2$                  & -90 dBm   \\ \hline
	\end{tabular}
\end{table}

\begin{figure}[!t]
	\centering
	\setlength{\abovecaptionskip}{0.cm}
	\includegraphics[width=11cm]{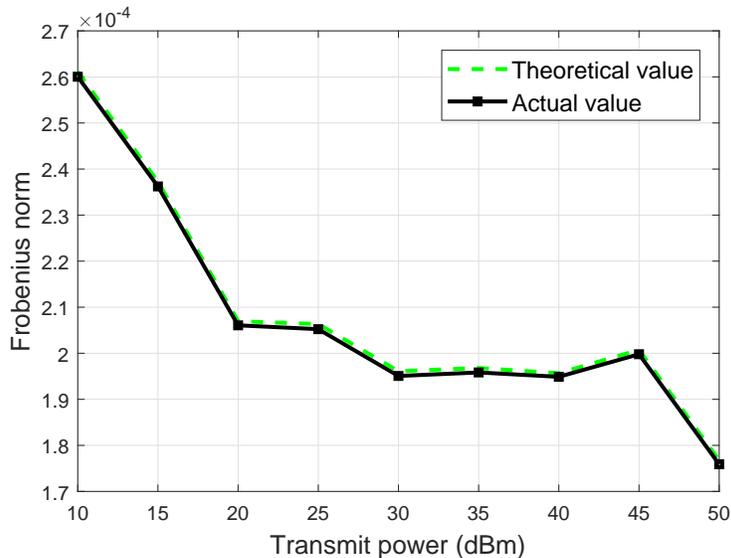}\\
	\caption{The differences between the Frobenius norm of the matrix ${\boldsymbol{\Sigma}}_{k}^{\left(1\right)}\left(i,j\right)$ in Theorem 1 and ${\beta _i} {\alpha _j}{\boldsymbol{\nu}}^{\textrm{H}} {{\boldsymbol{c}}^{ij}}$ obtained by our designed scheme.}\label{7}
\end{figure}

Fig. \ref{7} shows the differences between the Frobenius norm of the matrix ${\boldsymbol{\Sigma}}_{k}^{\left(1\right)}\left(i,j\right)$ in Theorem 1 and ${\beta _i} {\alpha _j}{\boldsymbol{\nu}}^{\textrm{H}} {{\boldsymbol{c}}^{ij}}$ obtained by our designed scheme. In this figure, we randomly select one user to compare its Frobenius norm of ${\beta _i} {\alpha _j}{\boldsymbol{\nu}}^{\textrm{H}} {{\boldsymbol{c}}^{ij}}$ with the F-norm of the theoretical value  of diagonal matrix ${\boldsymbol{\Sigma}}_{k}^{\left(1\right)}\left(i,j\right)$ in Theorem 1.  From this figure, we see that,  the Frobenius norm of ${\boldsymbol{\Sigma}}_{k}^{\left(1\right)}\left(i,j\right)$ and ${\beta _i} {\alpha _j}{\boldsymbol{\nu}}^{\textrm{H}} {{\boldsymbol{c}}^{ij}}$ are very close, which verifies the correctness of Theorem 1. From Fig. \ref{7}, we can also see that, the F-norm of ${\boldsymbol{\Sigma}}_{k}^{\left(1\right)}\left(i,j\right)$ and ${\beta _i} {\alpha _j}{\boldsymbol{\nu}}^{\textrm{H}} {{\boldsymbol{c}}^{ij}}$ fluctuates up and down with the change of the transmit power. This implies that the singular values of the mmWave effective channel are not affected by the 
transmit power.

\begin{figure}[!t]
	\centering
	\setlength{\abovecaptionskip}{0.cm}
	\includegraphics[width=11cm]{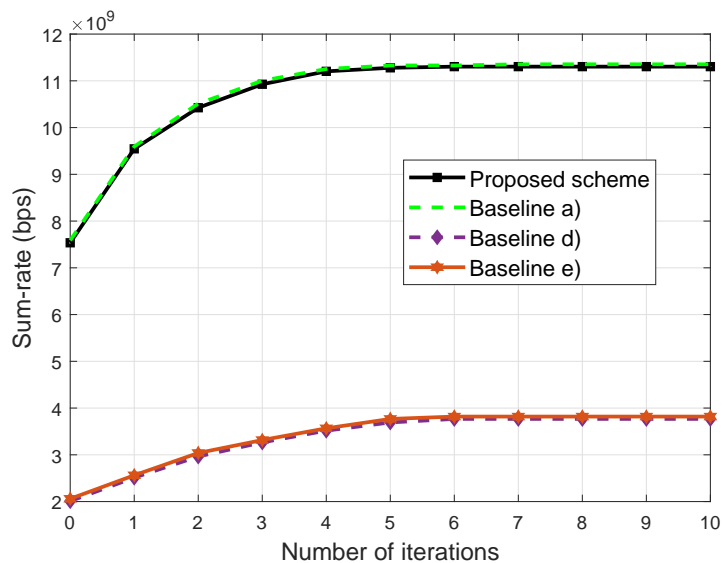}\\
	\caption{Convergence of the considered algorithms for sovling problem (\ref{PF17}).} \label{5}
\end{figure}

Fig. \ref{5}  shows the convergence  of the proposed algorithm to solve problem (\ref{PF17}). From Fig. \ref{5}, we observe that, the proposed algorithm  can achieve up to 2x and  2x gains in terms of the sum of all users' data rates compared to baselines d) and e). This is due to the fact that, our proposed BD method  can eliminate the inter-group interference.  From Fig. \ref{5}, we can also see that,  the number of iterations that the proposed algorithm needs to converge is similar to that of baseline a). This implies that the proposed algorithm that uses hybrid beamforming can reduce energy consumption without increasing the complexity of finding optimal solution for serving users. Fig. \ref{5} also shows that, the proposed scheme only needs seven iterations to solve problem (\ref{PF17}), which further verifies the quick convergence speed of the designed scheme.

\begin{figure}[!t]
	\centering
	\setlength{\abovecaptionskip}{0.cm}
	\includegraphics[width=11cm]{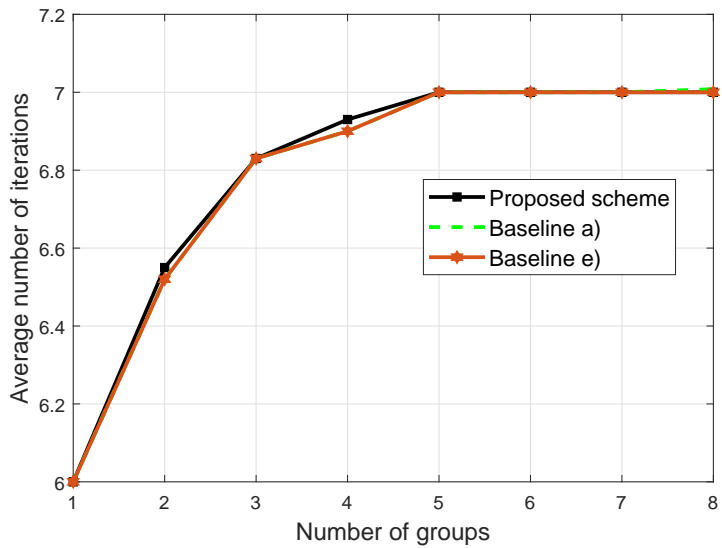}\\
	\caption{The average number of iterations versus number of groups.} \label{6}
\end{figure}

Fig. \ref{6} shows how the average number of iterations that the considered algorithms need to convergence changes as the number  of groups varies. From Fig. \ref{6}, we  see that, as the number of groups increases, the average number of iterations that the considered  algorithms need to converge increases. This is due to the fact that, as the number of groups increases, the considered algorithms require more iterations to find the optimal phase shifts of the IRS. As the number of groups continues to increase, the average number of iterations for convergence  remains constant. This is because the BS has enough user groups to    determine the phase shifts of the IRS.     From Fig. \ref{6}, we can also see that,  the  average number of iterations that the proposed algorithm needs to converge is similar to that of baselines a) and e). This implies that the proposed algorithm can reduce the hardware implementation complexity without increasing the complexity of finding optimal solution for serving users.

\begin{figure}[!t]
	\centering
	\setlength{\abovecaptionskip}{0.cm}
	\includegraphics[width=11cm]{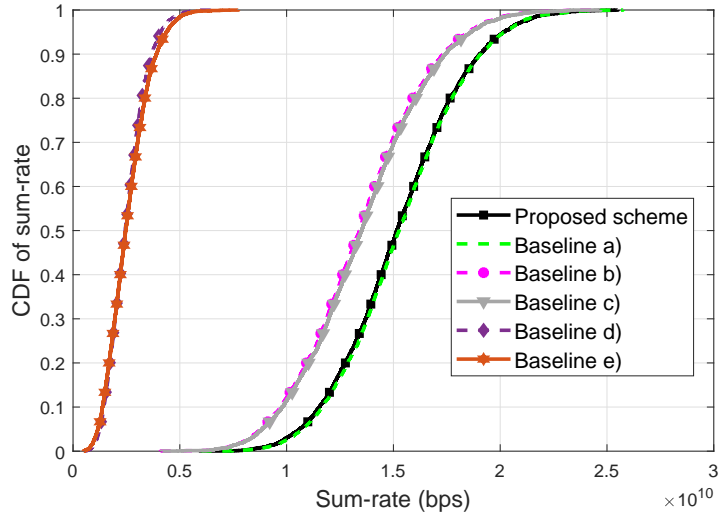}\\
	\caption{CDF of sum-rate.} \label{4}
\end{figure}

Fig. \ref{4} shows the cumulative distribution function (CDF) of the sum-rate of all users when the  transmit power of the BS is 40 dBm. 
From Fig. \ref{4}, we can see that the  proposed scheme
improves the CDF of up to 81.8\% and 78.33\% gains compared to baselines b) and c) when the sum of all users' data rates is 15 Gbps. This is because the phase shifts of the  proposed scheme are  optimized, resulting in high gain of the effective channel, which further facilitates the selection of the transmit beamforming matrices of the BS and the receive beamforming matrices of the users.

\begin{figure}[!t]
	\centering
	\setlength{\abovecaptionskip}{0.cm}
	\includegraphics[width=11cm]{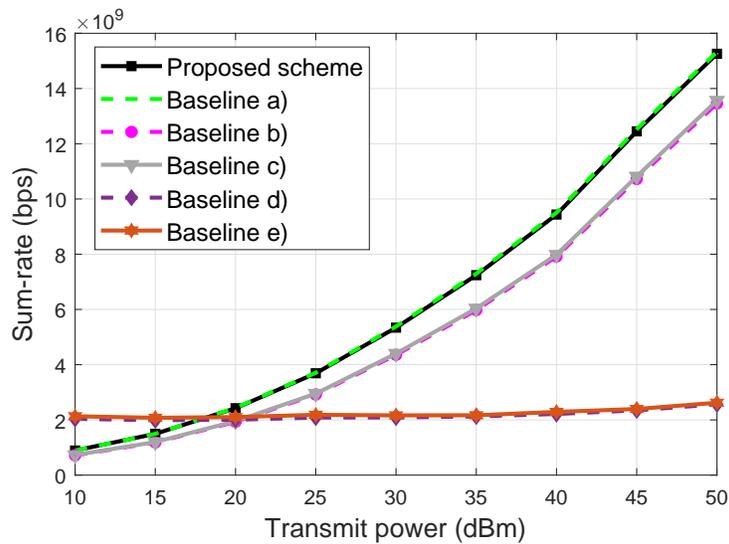}\\
	\caption{Sum-rate changes as the transmit power of the BS.} \label{1}
\end{figure}

Fig. \ref{1}  shows how the sum rate of all users changes as the transmit power of the BS varies. From Fig. \ref{1}, we  see that, the performance achieved by the proposed algorithm is similar to the optimal performance achieved by baseline a). This is because the proposed shceme can find  the optimal hybrid beamforming matrices to represent the fully digital matrices. Fig. \ref{1} also shows that compared to baselines b) and d),  the proposed scheme  can achieve up to 13\% and 5x gains in terms of the sum rate of all users when $P$=50 dBm and $M$=256. This is because the proposed scheme  optimizes the phase shifts of the IRS by a manifold method and eliminate the interference by the BD method. From Fig.~\ref{1}, we can also see that, as  the transmit power of the BS increases, the sum rate of baselines d) and e) remains unchanged. This is due to the fact that baselines d) and e) do not eliminate inter-group interference, which increases as the transmit power of the BS increases.
\begin{figure}[!t]
	\centering
	\includegraphics[width=11cm]{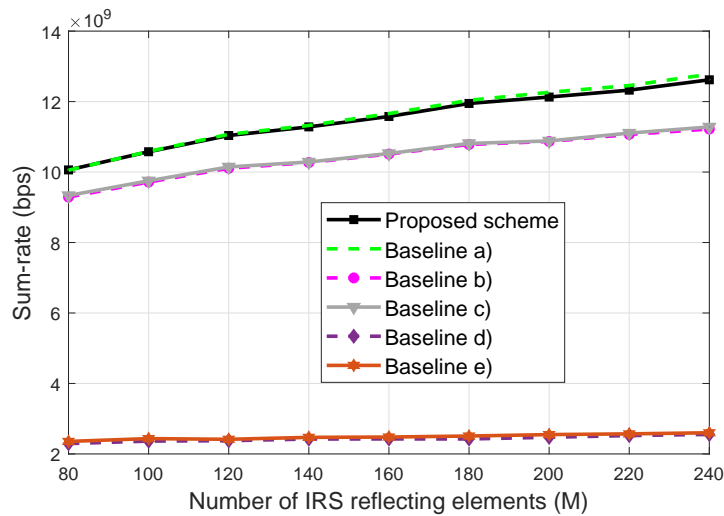}\\
	\caption{Sum-rate versus number of reflecting elements at the IRS.}\label{2}
\end{figure}

In Fig. \ref{2}, we show how the sum of all users' data rates changes as the number of reflecting elements at the IRS varies.  Fig. \ref{2}  shows that the proposed scheme  can achieve up to 13.3\% and  4x gains in terms of the sum of all users' data rates compared to baselines b) and d) when $M$=240. This is due to the fact that the proposed  scheme can align the angles of the cascaded channel and improve SINR. Fig. \ref{2} also shows that as the number of reflection elements of the IRS increases, the performance of baselines d) and e) is remains unchanged. This is because baselines d) and e) only eliminate the self-interference of each user without eliminating the inter-group interference.

\begin{figure}[!t]
	\centering
	\includegraphics[width=11cm]{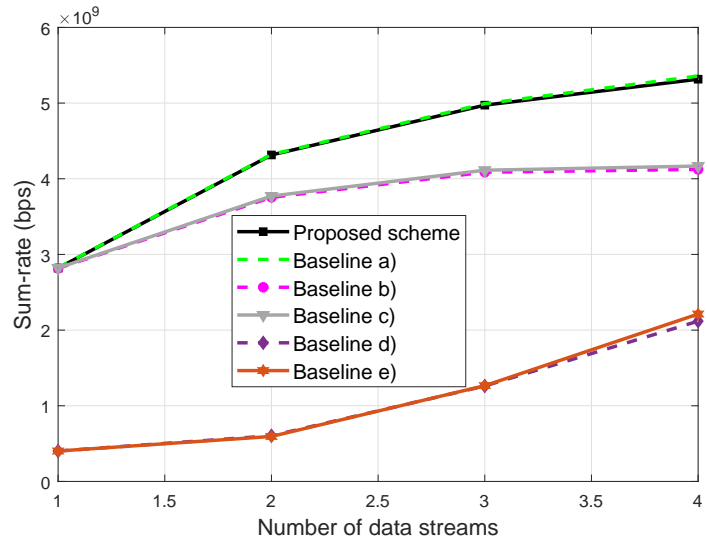}\\
	\caption{Sum-rate versus number of data streams.}\label{3}
\end{figure}

In Fig. \ref{3}, we show how the sum of all users' data rates changes as the number of data streams changes. From this figure, we can see that, as the number of data streams increases, the sum-rate of all considered algorithms increases. This is due to the fact that the desired signal power increases as the number of data streams increases. Fig. \ref{3} also shows that the proposed scheme  can achieve up to 28.6\% and  152.97\% gains in terms of the sum of all users' data rates compared to baselines b) and d) when the number of data streams is 4. This is because our proposed BD method can eliminate the inter-group interference and the self-interference of each user, and the manifold method  can find the suitable IRS phase shifts.

\begin{figure}[!t]
	\centering
	\includegraphics[width=11cm]{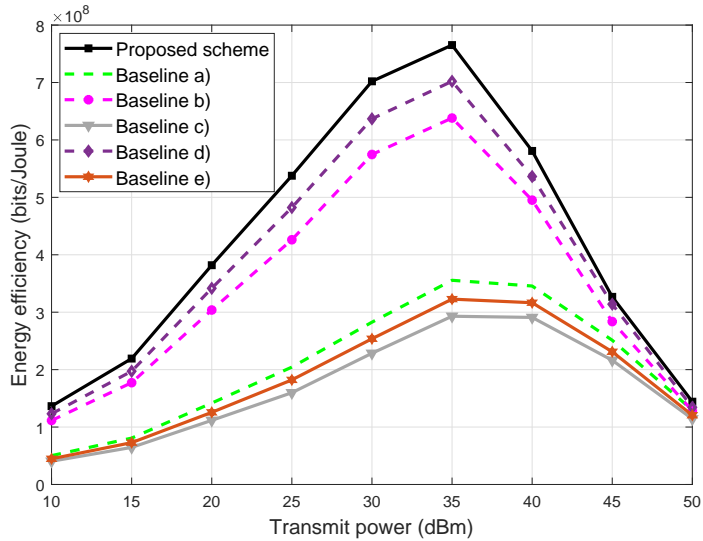}\\
	\caption{Energy efficiency versus the transmit power of the BS.}\label{9}
\end{figure}

Fig. \ref{9}  shows how the energy efficiency changes as the transmission power of the BS varies. In this figure, the energy efficiency is defined as the ratio of the sum-rate to the total power consumption of the system. The total power consumption of the system includes the transmission power at the BS, the hardware static power consumption at the BS, and the hardware static power consumption of the each reflecting element at the IRS. From Fig. \ref{9}, we can see that, the proposed scheme can achieve up to 19.92\%  gain in terms of energy efficiency compared to baseline b). This is due to the fact that the proposed algorithm optimizes the  phase shift to align the angle of the path from the BS to the users. From Fig. \ref{9}, we can also see that the proposed scheme can achieve up to 9\% gain in terms of energy efficiency compared to baseline d). This is because the proposed BD method eliminates both the inter-group interference and the self-interference, while baseline d) only eliminates the self-interference of each user. Fig. 9 also shows that the energy efficiency increases when the transmit power of the BS is less than 35 dBm. However, when the transmit power of the BS is higher than 35 dBm, the energy efficiency decreases. This is due
to the fact that, as the transmit power of the BS increases,
the total power consumption of the system   increases.

\section{Conclusions}
In this paper, we have developed a novel framework  for an IRS-assisted mmWave multigroup multicast MIMO communication system.  The transmit beamforming matrices of the BS, the  receive beamforming matrices of the users, and the phase shifts of the IRS were jointly optimized to maximize the sum rate of all users. We have used a BD method to represent the beamforming matrices of the BS and the users in terms of  the IRS phase shifts. Then, we have transformed the original problem to a problem that only needs to optimize the IRS phase shifts. The transformed problem is solved by a manifold method. Simulation results show that the proposed scheme can achieve significant performance gains compared to baselines.

\appendices
\section{}\label{Proof of thm1}
To prove Theorem \ref{thm1}, we first define the effective channel ${\boldsymbol{H}}_{k}$ as done in \cite{9234098}:   \vspace{-.8em}
\begin{equation}\label{PF11}  \vspace{-.5em}
{\boldsymbol{H}}_{k}  =  {G_\textrm{t}} {G_\textrm{r}} \boldsymbol{H}^\textrm{R}_{k} \boldsymbol{\Phi} \boldsymbol{H}^\textrm{B}=\boldsymbol{A}_{k} \boldsymbol{D}_{k} \left(\boldsymbol{A}\right)^{\textrm{H}},
\end{equation}
where $\boldsymbol{A}_{k}=\left[\boldsymbol{a}\left(r_{1,k}^{\textrm{A}}\right),\ldots,\boldsymbol{a}\left(r_{L,k}^{\textrm{A}}\right)\right]$ is a array response matrix of user $k$, $\boldsymbol{A}=\left[\boldsymbol{a}\left(r_1^{\textrm{D}}\right),\ldots,\boldsymbol{a}\left(r_Y^{\textrm{D}}\right)\right]$ is a array response matrix of the BS, and $\boldsymbol{D}_{k}$ is an $Y \times L$ matrix with element $\boldsymbol{D}_{k}\left(i,j\right)={\beta _i} {\alpha _j}  {d_{ij}}$, where ${d_{ij}}$ can be given by 
\begin{equation}\label{PF11+}
\begin{split}
{d_{ij}} & = \left(\boldsymbol{a}\left({\theta}_{i}^{{\textrm{D}}},{\eta}_{i}^{{\textrm{D}}}\right)\right)^{{\textrm{H}}}    \boldsymbol{\Phi}   \boldsymbol{a}\left({\theta}_{j}^{{\textrm{A}}},{\eta}_{j}^{{\textrm{A}}}\right)  \\
& ={\boldsymbol{\nu}}^{\textrm{H}}     \left(\left(\boldsymbol{a}\left({\theta}_{i}^{{\textrm{D}}},{\eta}_{i}^{{\textrm{D}}}\right)\right)^{{\textrm{*}}}    \circ    \boldsymbol{a}\left({\theta}_{j}^{{\textrm{A}}},{\eta}_{j}^{{\textrm{A}}}\right) \right)\\
& = {\boldsymbol{\nu}}^{\textrm{H}} {{\boldsymbol{c}}^{ij}}.
\end{split}
\end{equation}

Given the effective  channel ${\boldsymbol{H}}_{k}$, we define  $\tilde{\boldsymbol{H}}_{h}$ as 
\begin{equation}\label{PF12}
\begin{split}
&\tilde{\boldsymbol{H}}_{h}  =\left[ {\boldsymbol{H}}_1,\ldots,{\boldsymbol{H}}_{h-1},{\boldsymbol{H}}_{h+1},\ldots,{\boldsymbol{H}}_H \right]^{\textrm{T}},  \\
& =  \quad~~~~ \quad \quad \left[ \begin{matrix} \boldsymbol{A}_{1}{\boldsymbol{D}_{1}}{\left(\boldsymbol{A}\right)^{{\textrm{H}}}}&\\\boldsymbol{A}_{2}{\boldsymbol{D}_{2}}{\left(\boldsymbol{A}\right)^{{\textrm{H}}}}&\\\vdots&\\\boldsymbol{A}_{K}{\boldsymbol{D}_{K}}{\left(\boldsymbol{A}\right)^{{\textrm{H}}}}\end{matrix} \!\!\!\!\! \right],\\
& =   \left [ \begin{matrix} \boldsymbol{A}_{1}&\cdots&0&\\\vdots&\ddots&\vdots&\\0&\cdots&\boldsymbol{A}_{K} \end{matrix} \!\!\!\!\! \right ]  \left [ \begin{matrix} \boldsymbol{D}_{1}&\cdots&0&\\\vdots&\ddots&\vdots&\\0&\cdots&\boldsymbol{D}_{K} \end{matrix} \!\!\!\!\! \right ]  \left [ \begin{matrix} \left(\boldsymbol{A}\right)^{{\textrm{H}}}\\\vdots\\\left(\boldsymbol{A}\right)^{{\textrm{H}}} \end{matrix} \! \right ],\\
& = \left [ \begin{matrix} \boldsymbol{A}_{1}&\cdots&0&\\\vdots&\ddots&\vdots&\\0&\cdots&\boldsymbol{A}_{K} \end{matrix} \!\!\!\!\! \right ] \boldsymbol {P} \left [ \begin{matrix} \tilde{\boldsymbol{\Sigma}}&\boldsymbol {0}&\\\boldsymbol {0}&\boldsymbol {0}& \end{matrix} \!\!\!\!\! \right ] \boldsymbol {Q} \left [ \begin{matrix} \left(\boldsymbol{A}\right)^{{\textrm{H}}}\\\vdots\\\left(\boldsymbol{A}\right)^{{\textrm{H}}} \end{matrix} \! \right ].
\end{split}
\end{equation}
We define $\boldsymbol {Q} \left [ \begin{matrix} \left(\boldsymbol{A}\right)^{{\textrm{H}}}\\\vdots\\\left(\boldsymbol{A}\right)^{{\textrm{H}}}\end{matrix} \! \right ]$ as ${\boldsymbol{Z}}$, hence, $\tilde{\boldsymbol{V}}_{h}^{\left(0\right)}=\left [\boldsymbol{z}_{\left(K-{\left | \mathcal H_h \right |}\right){\zeta}+1}; \ldots; \boldsymbol{z}_{K{\zeta}}  \right ]$, with $\boldsymbol{z}_{\left(K-{\left | \mathcal H_h \right |}\right){\zeta}+1}$ being row ${\left(K-{\left | \mathcal H_h \right |}\right){\zeta}+1}$ of matrix ${\boldsymbol{Z}}$.      Based on (\ref{PF11}), ${\boldsymbol{H}}_{k} \tilde{\boldsymbol{V}}_{h}^{\left(0\right)}$ is given by       
\begin{equation}\label{PF13}
{\boldsymbol{H}}_{k} \tilde{\boldsymbol{V}}_{h}^{\left(0\right)}= \boldsymbol{A}_{k} \boldsymbol{D}_{k} \left(\boldsymbol{A}\right)^{{\textrm{H}}} \left [\boldsymbol{z}_{\left(K-{\left | \mathcal H_h \right |}\right){\zeta}+1}; \ldots; \boldsymbol{z}_{K{\zeta}}  \right ].
\end{equation}
For ULA with $N$ antennas, the column vectors of $\boldsymbol{A}_{k}$ and row vectors of $\left(\boldsymbol{A}\right)^{{\textrm{H}}} \left [\boldsymbol{z}_{\left(K-{\left | \mathcal H_h \right |}\right){\zeta}+1}; \ldots; \boldsymbol{z}_{K{\zeta}}  \right ]$ can form orthonormal sets\cite{9234098}. Hence, ${\boldsymbol{H}}_{k} \tilde{\boldsymbol{V}}_{h}^{\left(0\right)}=\boldsymbol{A}_{k} \boldsymbol{D}_{k} \left(\boldsymbol{A}\right)^{{\textrm{H}}} \left [\boldsymbol{z}_{\left(K-{\left | \mathcal H_h \right |}\right){\zeta}+1}; \ldots; \boldsymbol{z}_{K{\zeta}}  \right ]$ can be considered as an approximation of the truncated SVD of ${\boldsymbol{H}}_{k} \tilde{\boldsymbol{V}}_{h}^{\left(0\right)}$, and $\boldsymbol{D}_{k}$ can represent ${\boldsymbol{\Sigma}}_{k}^{\left(1\right)}$.


\bibliographystyle{IEEEtran}
\bibliography{IEEEabrv,MMM}
\end{document}